%Paper: astro-ph/9306023
%From: hst!philf@physics.att.com
%Date: Wed, 23 Jun 93 17:51:35 EDT

%Five postscript figures are included at the end

\magnification=\magstep1
\hsize = 6.5 true in\vsize = 9.0 true in
\baselineskip = 12 pt
\overfullrule=0pt
\def\titrule{\vrule height0.4pt depth0.0pt width4.0cm}
\footline={\hfil}
\headline={\ifnum\pageno=1 \hfil \else\hfil\folio \fi}

\font\fbell=cmss10
\font\sevenrm=cmr7
\font\sevenit=cmmi7
\font\sevensy=cmsy7
\def\sevenpoint{\def\rm{\fam0\sevenrm}
\textfont0=\sevenrm
\textfont1=\sevenit
\textfont2=\sevensy
\def\it{\fam\itfam\sevenit}
\textfont\itfam=\sevenit
}

\font\eightrm=cmr8
\font\eightit=cmmi8
\font\eightsy=cmsy8
\font\eightbf=cmbx8
\def\eightpoint{\def\rm{\fam0\eightrm}
\textfont0=\eightrm
\textfont1=\eightit
\textfont2=\eightsy
\def\it{\fam\itfam\eightit}
\textfont\itfam=\eightit
}

\font\ninerm=cmr9
\font\nineit=cmmi9
\font\ninesy=cmsy9
\def\ninepoint{\def\rm{\fam0\ninerm}
\textfont0=\ninerm
\textfont1=\nineit
\textfont2=\ninesy
\def\it{\fam\itfam\nineit}
\textfont\itfam=\nineit
}

\newcount\tableone \tableone=0
\newcount\fignum \fignum=0
\newcount\fignumc \fignumc=0
\newcount\equa \equa=0
\newcount\itemm \itemm=0
\font\fchap=cmbx12
\font\ftext=cmr10
\font\fsmall=cmr7

\def\taone{\global\advance\tableone by 1 \the\tableone}
\def\taonen{\the\tableone{\ (cont.)}}
\def\itemno{\global\advance\itemno by 1 \the\itemoone}
\def\takeep{\the\tableone}
\def\fig{\global\advance\fignum by 1 \the\fignum}
\def\figcc{\global\advance\fignumc by 1 \the\fignumc}
\def\equat{\global\advance\equa by 1 \the\equa}
\def\itemno{\global\advance\itemm by 1 \the\itemm}
\def\itemn{\item{\itemno)}}

\newdimen\digitwidth\setbox0=\hbox{\rm0}\digitwidth=\wd0
\catcode`~=\active\def~{\kern\digitwidth}
\catcode`!=\active\def!{\kern 0.7\digitwidth}
%\catcode`|=\active\def|{\kern 1.4\digitwidth}

\edef\innernewbox{\noexpand\newbox}
\def\sbl{
\innernewbox\tablebox
\centerline { \box\tablebox}
\setbox\tablebox = \vbox \bgroup\offinterlineskip
\halign}

\def\tablerule{\noalign{\hrule}}
\def\pmm#1#2{{\kern 1.4\digitwidth}^{#1}_{#2}}
\def\Deg{${}^\circ$\llap{.}}
\def\Min{${}^{\prime}$\llap{.}}
\def\Sec{${}^{\prime\prime}$\llap{.}}
\def\deg{${}^\circ$}
\def\min{${}^{\prime}${ }}
\def\sec{${}^{\prime\prime}${ }}
\def\sol{$_\odot$}
\def\et{{\it et al.\ }}
\def\Rc2{{\it R}}
\def\vave{\overline {\hbox{v}}}
\def\figur#1{
\pageinsert
%\epsfxsize=\hsize \epsfbox{#1}
%\special{psfile=#1 hoffset=-589 hscale=83 vscale=83}
\vfil
\noindent
\centerline{Fischer et al. -- Fig. \fig}
\endinsert}
\def\kms{km s$^{-1}$}

\def\tabhl#1#2{{\ftext
\centerline{Table #1}
\centerline{#2}}
\innernewbox\tablebox
\centerline { \box\tablebox}
\setbox\tablebox = \vbox \bgroup\offinterlineskip
\halign\bgroup}

\def\tabhlr#1#2{
\innernewbox\tablebox
%\centerline
{\box\tablebox}
\setbox\tablebox = \vbox \bgroup\offinterlineskip
{\ftext
\centerline{#1}
\smallskip
\centerline{#2}
\medskip
}
\halign\bgroup}

\edef\inskip{\noexpand\newskip}
\def\splr{
\noalign{}
\tablerule
\noalign{\vskip0.1truecm}
\tablerule\egroup\egroup
\skip3=\wd\tablebox
\skip4=\ht\tablebox
\skip1=0pt
\advance\skip1 by \skip3
\advance\skip1 by -662.2pt
\divide\skip1 by 2
\skip2=0pt
\advance\skip2 by \skip4
\advance\skip2 by 59.5pt
\advance\skip2 by \skip1
\vskip-\skip2
\rotr\tablebox
}

\long\def\tabh#1#2{{\ftext
\centerline{Table #1}
\centerline{#2}}
\innernewbox\tablebox
\setbox\tablebox = \vbox \bgroup
\halign\bgroup}

\long\def\tabhr#1#2{\innernewbox\tablebox
\setbox\tablebox = \vbox \bgroup
{\ftext\hsize=9truein
\centerline{Table #1}
\smallskip
\centerline{#2}
\medskip
\hsize=6.5truein}
\halign\bgroup}

\def\sp{\noalign{\vskip.2truecm}
\tablerule
\noalign{\vskip0.1truecm}
\tablerule\egroup\egroup
\centerline{\box\tablebox}}
\def\spc{\noalign{\vskip.2truecm}
\tablerule\egroup\egroup
\centerline{\box\tablebox}}

\def\spl{
\noalign{}
\tablerule
\noalign{\vskip0.1truecm}
\tablerule\egroup\egroup
\centerline {\box\tablebox}}

\def\splc{
\noalign{}
\tablerule\egroup\egroup
\centerline {\box\tablebox}}

\def\topper{
\noalign{\vskip0.3truecm}
\tablerule
\noalign{\vskip.1truecm}
\tablerule
\noalign{\vskip.1truecm}
}
\def\topperc{
\noalign{\vskip0.3truecm}
\tablerule
\noalign{\vskip.1truecm}
}
\def\topperl{
\noalign{\vskip0.3truecm}
\tablerule
\noalign{\vskip.1truecm}
\tablerule
}
\def\spacer{
\noalign{\vskip0.2truecm}
\tablerule
\noalign{\vskip0.1truecm}
}

\xdef\amper{&}
\newcount\n
\def\hspa#1{}
\def\tcoli{## \hfil}
\def\tcol#1{\hskip#1cm\hfil ## \hfil}
\def\tcolil#1#2{## \hfil\vrule height #1pt depth #2pt}
\def\tcoll#1#2#3{\hskip#1cm\hfil ## \hfil\vrule height #2pt depth #3pt}
\def\vbs#1#2{\hfil\vrule height #1pt depth #2pt}
\def\vvbs#1#2{\vrule height #1pt depth #2pt}
\def\mcol#1#2{\n=0
\loop\ifnum\n<#2 \advance\n by1
\hskip#1cm\hfil ## \hfil \noexpand\amper \repeat}
\def\figc{\noindent Fig. \figcc --\ }

\global\def\sectname{}
\newcount\sectnum \sectnum=0
\newcount\subnum \subnum=0
\def\sect#1{\medskip\penalty-8000
\global\advance\sectnum by 1 \centerline{{\the\sectnum{.} \hskip1em #1}}
\def\sectname{#1}
\subnum=0
\medskip
\penalty8000
}
\def\subsect#1{\medskip\penalty-5000
\global\advance\subnum by 1 \centerline{\it {{\the\sectnum{.}\the\subnum
\hskip1em #1}}}
\medskip
\penalty5000
}
\def\ref#1#2#3#4{#1, {\it #2}, {#3}, #4.
\smallskip}
\def\apj{ApJ}
\def\apjs{ApJS}
\def\aj{AJ}
\def\aa{A\&{A}}
\def\aas{A\&{A}S}
\def\pasp{PASP}
\def\mnras{MNRAS}
\def\fullref#1{#1. \smallskip}
\pageinsert
\baselineskip = 15pt
%\hskip2.1truein{\epsfbox{../letter/attlogo.epsi}}
%\hskip-1.95truecm {\fbell Bell Laboratories}
\medskip
\hrule
\bigskip\bigskip
\centerline{\bf DYNAMICS OF THE GLOBULAR CLUSTER NGC 362}
\bigskip\bigskip
\centerline{PHILIPPE FISCHER$^1$}
\centerline{Department of Physics and Astronomy, McMaster University, Hamilton,
Ontario L8S
4M1, Canada}
\centerline{and}
\centerline{AT\&T Bell Laboratories, 600 Mountain Ave., 1D-316, Murray Hill, NJ
07974}
\centerline{Email: philf@physics.att.com}
\centerline{}
\centerline{DOUGLAS L. WELCH$^{1}$}
\centerline{Department of Physics and Astronomy, McMaster University, Hamilton,
Ontario L8S
4M1, Canada}
\centerline{ }
\centerline{MARIO~MATEO$^{2}$}
\centerline{The Observatories of the Carnegie Institute of Washington}
\centerline{813 Santa Barbara Street, Pasadena, CA 91101}
\centerline{}
\centerline{PATRICK C\^OT\'E}
\centerline{Department of Physics and Astronomy, McMaster University, Hamilton,
Ontario L8S
4M1, Canada}
\centerline{ }
\vskip 0.50 truein
\newdimen\addwidth\setbox0=\hbox{Address for proofs: }\addwidth=\wd0
\vskip 0.25 truein
\leftline{Address for proofs:}
\leftline{Philippe Fischer}
\leftline{AT\&T Bell Laboratories}
\leftline{1D-316}
\leftline{600 Mountain Ave.}
\leftline{Murray Hill, NJ 07974}
\leftline{Email: philf@physics.att.com}
\medskip
\centerline{To appear in the October 1993 Astronomical Journal}
\vfil
\line{ \titrule \hfill}
\noindent
$^1$ Guest Investigator, Mount Wilson and Las Campanas Observatories,
which are operated by the Carnegie Institution of Washington.
\centerline{ }
\noindent
$^2$ Hubble Fellow.
\baselineskip = 20 pt
\endinsert
\centerline{\bf Abstract}
\medskip

In this paper we have examined the internal dynamics of the globular cluster
NGC 362 using a combination of V band CCD images and echelle spectra of
member red giants. A V band surface brightness profile (SBP) was constructed
from the CCD images, and, after it was determined that the cluster is not
post core-collapse, fit with single- and multi-mass King-Michie (KM) models.
We found that for small values of the mass function slope, $x$, anisotropic
models were favored while for steeper mass functions isotropic orbits
provided superior fits.  The total cluster luminosity is 1.70 $\pm 0.1
\times 10^5$ L$_V$\sol [assumes (m--M)$_0$ = 14.77].  A total of 285 stellar
spectra were obtained of 215 stars for radial velocity determinations. Three
stars were obvious non-members and four showed strong evidence for radial
velocity variations; these latter stars are probably members of binary
systems with periods less than a few years. The true cluster binary fraction
was determined from simulations to be 0.15 for circular orbits or 0.27 for
orbits with a distribution function $f(e) = e$ ($e$ is eccentricity).  This
relatively high binary detection frequency may indicate that NGC 362 is
overabundant in binaries compared to other clusters. The 208 remaining stars
showed no sign of rotation and had kinematics which were incompatible with
KM models having isotropic orbits and luminosity profiles consistent with
the SBP. Therefore, the best agreement with both the kinematic data and the
SBP were for shallow mass functions $x = 0.0 - 0.5$ and intermediate amounts
of anisotropy in the velocity dispersion tensor. In this best-fit range, the
derived cluster mass is M = $2.5 - 3.5 \times 10^5$ M\sol\ for a global
mass-to-light ratio of M/L$_V$ = 1.5 - 2.0 M\sol/L$_V$\sol. This low value
for $x$ is in disagreement with the correlation between $x$ and the height
above the Galactic disk seen for a sample of other clusters. The results are
also different from the sharp turn-up in the low mass end of mass functions
derived from some deep luminosity functions of three other globular clusters.

\vfill\eject
\sect{Introduction}

Kinematic studies are a potentially powerful means of learning about the
present dynamical state and stellar content of globular clusters.  They
can lead to important conclusions regarding the history of the globular
cluster system and the low-end of the stellar mass function which impact on
star formation and the formation and evolution of the Galactic halo.

The exact nature of the low mass end of the halo mass function is highly
uncertain. Obtaining this information through observations of individual
field halo stars is very difficult as one is faced with contamination from
faint galaxies, uncertain distances and small number statistics. One can
overcome these problems by studying the luminosity function of halo stars
which are members of globular clusters. Unfortunately these studies are
generally confined to small regions at large projected radii from the
cluster centers and, with a very few noteworthy exceptions, sample only a
small range in mass (i.e.  0.5 - 0.8 M\sol).  Other problems include
crowding, background contamination and corrections to the local mass
function for mass segregation.

Kinematic measurements along with dynamical modeling can be used to overcome
some of the problems in the luminosity function approach, but also possess
some intrinsic shortcomings. The Achilles heel of this method is the
problem of uniqueness which occurs whenever one is dealing with a system
where mass does not follow light (galaxy, and galaxy clusters share this
problem). In such cases, it is possible to obtain a large range of models
possessing different dynamical parameters (i.e. mass, orbital anisotropy)
which are similarly consistent with the observational data. The approach
that one makes, therefore, is to restrict the parameter space to a
``reasonable'' subset of the possible range.

Until recently, obtaining large samples of high precision radial velocities
has been a very time-consuming task. Thus, there are very few kinematically
well-studied clusters (i.e. possessing more than 100 or so radial
velocities). Among the better-sampled clusters are M3 (Gunn \& Griffin 1979),
Omega Cen and 47 Tuc (Meylan \& Mayor 1986), M13 (Lupton, Gunn, \& Griffin
1987), M15 (Peterson \et 1989), NGC 6397 (Meylan \& Mayor 1991), NGC 3201
(C\^ot\'e \et 1993), with several large data sets in existence yet to be
published.

In this paper we present a dynamical analysis of NGC 362. This cluster is an
ideal candidate for a radial velocity study because of its high systemic
radial velocity (223.5 \kms) which enables one to unambiguously remove
contaminating field stars. Furthermore, it is a particularly difficult
cluster for which to obtain a luminosity function because it is projected on
to the halo of the Small Magellanic cloud. In \S 2.1 and 2.2 we discuss the
CCD imaging and echelle spectroscopy observations and reductions. In \S 3.1
and 3.2 we describe the ``reasonable'' dynamical parameter subspace we have
chosen for this analysis, that is the popular King-Michie models which
invoke mass segregation, tidal truncation and velocity anisotropy with a
minimum of free parameters. In \S 3.3 we summarize the results of the
modeling, including a comparison with some previous luminosity and mass
function studies. \S 3.4 is a justification of the use of the King-Michie
models based on the NGC 362 relaxation timescales.

\sect{Observations and Reductions}

\subsect{Surface Photometry}

Four V CCD frames of NGC 362 were obtained at the Las Campanas Observatory
(LCO) 1.0 m telescope on 1991 January 22.  The TEK2 1024$^2$ chip was used
(readout noise = 7 e$^-$, gain = 2 e$^-$/ADU, and angular scale =
0.61$^{\prime\prime}$ px$^{-1}$). The exposure times were 30 seconds for
each image. The frames were positioned with the cluster in one corner of
each image such that full azimuthal coverage was obtained.

The coordinates of the cluster center were adopted from Shawl \& White
(1986).  Astrometric zero-points for each frame were determined through the
cross-identification of 15 stars from the Guide Star Catalog (Lasker
\et 1990, Russell \et 1990, and Jenkner \et 1990). The RMS uncertainty in
the position of a selected Guide Star was 0.5 arcsec.

Surface photometry was performed in a manner similar to Djorgovski (1988).
The frames were broken up into a series of concentric circular annuli
centered on the cluster.  The annuli were further divided into eight
azimuthal sectors. The average pixel brightness was determined for each
sector in a given annulus and the {\it median} of the eight separate
measurements was taken as the representative brightness at the area-weighted
average projected radius of the annulus (i.e., the mean radius of all the
pixels within the annulus which is approximately equal to the geometric
mean). The standard error of the median of the eight sectors was adopted as
the photometric uncertainty.

A background level (a combination of SMC stars, sky light and Galactic
foreground stars) was estimated from regions at large projected distances
from the cluster.  We found that the surface brightness profile (SBP) tended
to level out beyond 21 pc (we have adopted a distance to the cluster of 9
kpc, see below). By ``leveling out'' we don't necessarily mean that the
cluster light does not extend beyond this point but simply that fluctuations
in the background dominate to such an extent that it is no longer possible to
observe the profile declining in intensity.  Therefore, it was this region
with a projected radius of 21.0 $\le$ R (pc) $\le$ 33.0 that was used for
the background determinations. The background value was $195.7 \pm 1.6$
L$_V$\sol\ pc$^{-2}$.

The background-subtracted surface photometry data is presented in Table 1
[assuming a cluster distance of 9 kpc (see below), M$_{V\odot} = 4.83$
Mihalas and Binney 1981, p. 60]. Column 1 is the area-weighted angular
radius, column 2 is the projected area-weighted radius, column 3 is the V
luminosity surface density, and columns 4 and 5 are the inner and outer
radii of the annuli, respectively.  Calibration was accomplished by matching
up 10 on-frame standards from Alcaino \et (1988).  The RMS scatter was 0.055
mag.

\subsect{Radial Velocities}

Spectra of red giants and horizontal branch stars in the region surrounding
NGC 362 were obtained during two runs: 1989 December 7 -- 14, and 1991
December 15 and 17, using the photon-counting echelle spectrograph on the
2.5m Dupont reflector, designed and built by Steve Shectman. The observation
and reduction procedures have been discussed extensively in Welch \et (1991)
and remain largely unchanged for these data. Briefly, the observing procedure
consisted of exposures with integration times of 100 - 500s and Th-Ar arcs
approximately every 45 minutes. A representative LCO spectrum is shown in
Fig. 2 of C\^ot\'e \et (1991).  The reduction utilizes the IRAF ECHELLE and
RV packages (Tody 1986) to obtain both velocities and velocity uncertainties
according to the prescription of Tonry \& Davis (1979).  The velocity
zero-point is tied to the IAU velocity standard 33 Sex as described in
Fischer \et 1992 and is accurate to better than 2 \kms.

A total of 285 stellar spectra were obtained. There were 215 distinct stars
observed; fifty were observed twice, one observed three times, one observed
ten times, and one was observed eleven times. The heliocentric radial
velocities are listed in Table 2, where the columns from left to right are:
the star designation, the equinox J2000.0 right ascension and declination,
the derived projected radius in arcsec and position angle in degrees, the
heliocentric Julian Date, the observed radial velocity and uncertainty, the
mean radial velocity and uncertainty for stars with multiple observations,
the $\chi^2$ and number of degrees of freedom for stars with multiple
observations, and comments. The designations are taken from Harris (1982)
when possible. The adopted absolute distance modulus (see \S 3.1) of
(m--M)$_0$ = 14.77 mag corresponds to 0.0436 pc/arcsec.

As pointed out above, our complete data set includes 53 stars having radial
velocities corresponding to two different epochs (separated by roughly two
years). Such a timespan should be sufficient for identifying cluster
binaries, and, indeed, four of these 53 stars (or 7.5\%) show velocity
differences greater than 6 km s$^{-1}$. Motions in the atmospheres of stars
near the tip of the RGB rarely exceed about 5 km s$^{-1}$ (Mayor {\it et
al.} 1984). We therefore suspect these discrepancies arise from the effects
of a companion since an inspection of the catalog of (known) NGC 362
variables (Sawyer-Hogg 1973) revealed only one such star in our sample
(H1204=V11).  If we insist on restricting ourselves only to those stars
showing variations greater than 8 km s$^{-1}$, our number of detected
binaries drops to three (or 5.7\%).  In either case, the NGC 362
binary detection frequency appears significantly higher than that seen in
other globular clusters (Hut {\it et al.} 1992).

In order to convert these detection frequencies into true binary fractions,
we have carried out a series of Monte Carlo simulations similar to those of
Pryor {\it et al.} (1988). We have generated 10000 simulated radial
velocities pairs (separated by two years) for a grid of hypothetical
clusters having binary fractions, $x_b$ = 0.00, 0.05, 0.10, 0.15, 0.20,
0.25, 0.30, 0.35, 0.40, 0.45 and 0.50. Unlike the models of Pryor {\it et
al.} (1988), which include a second free parameter, r$_{min}$, (the minimum
orbital separation which avoids mass transfer) we have randomly assigned an
r$_{min}$ to each binary by choosing values in the range 0.05 -- 0.14 AU.
The vast majority of our program objects uniformly lie in the range 14.0
$\le$ V $\le$ 15.5 which, according to the [Fe/H] = -1.03, [O/Fe] = 0.5, T =
13 Gyr isochrone of Bergbusch and VandenBerg (1992) (see \S 3.1),
corresponds to the above primary radii. Simulations have been carried out
for the two cases of (1) purely circular orbits and (2) a distribution of
orbital eccentricities given by $f(e)=e$ (Heggie 1975). The models use the
distributions of orbital periods and mass ratios described by Pryor {\it et
al.} (1988).  In all cases, we assume random orbital inclinations,
orientations, phases and observational error.

In Table 3 we show the fraction of stars expected to show velocity
variations greater than 6 or 8 km s$^{-1}$ based on these simulations. The
first column gives the assumed binary fraction while the second and third
columns record the expected fractions for both circular and eccentric
orbits, respectively. By comparing our observed detection frequency to the
model results, we see that the true NGC 362 binary fraction is roughly 0.15
for circular orbits or 0.27 for a Heggie (1975) distribution of orbital
eccentricities.  Of course, these numbers may be slightly enhanced by our
procedure of preferentially selecting for repeat measurement those stars
which showed higher than normal residual velocities.  Nevertheless, the
derived binary frequency for NGC 362 seems rather higher than that seen by
Hut {\it et al.} (1992). Clearly, further monitoring of these four stars
(H1348, H1419, H2205 and H2222) is in order, especially the luminous RGB
star H2205 which shows extreme velocity variability (but no known photometric
variability whatsoever (Alcaino 1976 and Harris 1982).

For the following dynamical analysis, we have therefore rejected seven stars
from the sample in Table 2. In addition to the four binaries mentioned
above, the RR Lyrae H1204 and the two (photometric and radial velocity)
non-members H1423 and H2113 have also been discarded. For the remaining
stars which have repeat measurements, the total $\chi^2$ is then 70.47 for
67 degrees of freedom which corresponds to P($\chi^2$) $\approx$ 0.35.
The analytical error estimates returned by RVXCOR therefore seem to
be reasonable (for this particular data set anyway) and we do not need to add a
so-called $``$jitter$"$ term which other investigators have found necessary
to account for possible motions in the atmospheres of these evolved stars.
It is possible that our error estimates are more generous and therefore take
into account this atmospheric jitter or simply that other studies, which
have generally relied on radial velocity scanners, have tended to
underestimate the velocity uncertainties.

The 208 remaining radial velocities are plotted in Fig. 1 versus projected
radius (top panel) and position angle (bottom panel). There is no obvious
indication of rotation in the velocity versus position angle graph which
should manifest itself as a sinusoidal trend if aligned close to equator-on.
An alternative way to test for rotation is to measure the difference in
median velocities for stars on either side of an imaginary axis as it is
stepped around the cluster center in 1\deg intervals.  Fig. 2 is a plot of
this and the solid line is the best-fitting sine curve having an amplitude
of 0.38 \kms. While there does seem to be a correlation in the data, we
found, based on 1000 simulations of the data assuming a non-rotating model,
that this amplitude is not significant. That is, one would expect this
amplitude or higher over 90\% of the time if the cluster was not rotating.
Therefore, we conclude that if rotation is present in NGC 362, it is too
small an effect compared to the velocity dispersion for us to detect (see \S
2.2). Furthermore, if rotation is present, it is not dynamically important,
meaning that the assumption of non-rotation and sphericity, consistent with
the small cluster ellipticity of $\epsilon = 0.99$ found by White \& Shawl
(1987), will not significantly bias our mass estimates. If the cluster is
being viewed pole-on or at a significant inclination, rotation would not be
observable even if present. However, the small amounts of flattening seen in
Milky Way globular clusters indicates that rotation is probably
dynamically insignificant in most of them.

Finally, Suntzeff \et (1986) reported lower precision velocities for four
stars in NGC 362: H1216 (242.3 km s$^{-1}$), H1441 (210.8), H2108 (225.8),
and H2431 (220.1). Our velocities for these stars are different by $-$15.0,
$+$18.5, $-$1.5, and $+$4.6 km s$^{-1}$, respectively, which suggests good
agreement in the velocity zeropoint considering that their estimated
uncertainty for an individual velocity was 15 km s$^{-1}$.

\sect{King-Michie Models}

We used maximum likelihood fits of single- and multi-mass King-Michie (KM)
models (King 1966, Michie 1963, Da Costa \& Freeman 1976 Gunn \& Griffin
1979) to the SBP in order to constrain the orbital distribution function and
obtain the cluster luminosity. Once this is accomplished, maximum
likelihood scaling is determined from the radial velocities yielding
cluster mass estimates.

\subsect{Surface Photometry}

The KM models require that the stellar mass spectrum be broken up into a
number of mass classes with mean stellar mass $m_i$ (see Table 4) each of
which is assumed to have an energy and angular momentum per unit mass ($E$
and $J$, respectively) distribution function given by $$f_i(E,J)
\propto e^{-[J/(2v_sr_a)]^2}[e^{-A_iE}-1],\eqno(\equat)$$ where $v_s$ is the
scale velocity, $W$ is the reduced gravitational potential, and r$_a$ is the
anisotropy radius beyond which stellar orbits become increasingly radial.
For a thorough discussion see Da Costa \& Freeman (1976) and Gunn \& Griffin
(1979). To fit the surface photometry, $W_0$ determines the shape of the
projected luminosity density distribution and two parameters, the scale
radius $r_s$, and the scale luminosity, are used convert this model density
to observed units, pc and L$_V$\sol\ pc$^{-2}$, respectively.

The $A_i$ are chosen to be proportional to the mean stellar mass of mass
class $i$, which approximates equipartition of energy in the cluster center.
Equipartition requires that $m_i\sigma_i^2 = m_j\sigma_j^2$ for all $i,j$, a
condition which is not met in the KM formulation due to the differing
effects of the tidal cut-off on the velocity dispersions of the different
mass classes (Pryor \et 1986). The effect mimics incomplete energy
equipartition; the high mass stars maintain too much kinetic energy relative
to the low-mass stars. This is qualitatively similar to the Fokker-Planck
evolutionary models of Inagaki and Saslaw (1985) in which the different mass
classes cease to interact after an initial period of energy exchange because
the lower mass stars are now located at larger radii than the high-mass
stars. The equipartition/mass segregation dynamics in globular clusters is a
complex issue which is certainly not fully addressed in the KM formulation.
However, by employing both single-mass models (no mass segregation, constant
mass-to-light ratios) and multi-mass models (which probably predict too much
mass segregation, see Pryor \et 1986), we explore the range in parameter
space within which the true cluster mass segregation state should lie.

To facilitate the comparison of the KM models, which yield mass density
profiles, to the SBPs, we need to invoke a stellar mass-luminosity
relationship. Bolte (1987) has published a CCD color-magnitude diagram (CMD)
of NGC 362. He applied isochrones from VandenBerg \& Bell (1985), which have
heavy element abundances which scale as the solar elements, to his
photometry.  Since that time, Dickens \et (1991) have carried out high
resolution echelle spectroscopy of eight member red giants obtaining [Fe/H]
$= -0.98 \pm 0.07$. They also determined that [(C+N+O)/H] $= -0.89 \pm 0.05$,
slightly overabundant compared to solar. We chose to use the stellar models
of Bergbusch \& VandenBerg (1992) having [Fe/H] = -1.03 and [O/Fe] = 0.5.
These models have the advantage of extending from the tip of the red giant
branch (RGB) down to 0.15 M\sol, but are probably overabundant in C,N, and
O. We show isochrones produced from this model for ages $\tau =$ 12, 13 and
14 Gyr, E(B--V) = 0.0 mag, and (m--M)$_0$ = 14.77 mag overlaid on the
main-sequence ridge line from Bolte (1987) in Fig. 3. We also show one of
Bolte's better fits having [M/H] = -0.79 and $\tau = 14$ Gyr, E(B-V) = 0.0
mag, and (m--M)$_0$ = 14.60 mag.  In the inset we have plotted V band
luminosity versus stellar mass.  The non-existent reddening that these
isochrones favor is somewhat inconsistent with the value of E(B-V) = 0.04 mag
that was found by both Burstein
\& Heiles (1982) from the foreground HI column density and Harris (1982) from
UBV photometry. A re-examination of Fig. 3 reveals that it would not be
possible to reconcile the above isochrones with anything more than a minute
amount of reddening, although, as Bolte points out, color uncertainties of a
few hundredths may exist in the photometry calibration and/or the model
color-temperature relationship.

As mentioned, the Bergbusch \& VandenBerg isochrones extend to the tip of
RGB.  However, the horizontal branch (HB) stars, despite their short
lifetimes, can contribute a significant fraction of the total cluster
luminosity and even more to the central luminosity. The CMD of Harris
(1992), demonstrates that the majority of the cluster HB stars are red and
range in V magnitude from 15.0 to 15.5 with a mean around 15.25
corresponding to a mean absolute magnitude of approximately M$_V = 0.5$ mag.
The main problem with attempting to incorporate the HB stars is estimating
their lifetimes. Fortunately, the HB evolution for stars corresponding to
the same model parameters as the Bergbusch \& VandenBerg isochrones is
described in Dorman (1992).  A typical lifetime from helium flash to core
helium exhaustion is 100.0 Myr and is fairly insensitive to the zero-age HB
mass.  By comparing this to the RGB timescales in VandenBerg (1992) we found
that 100.0 Myr is about the time it takes for a star to evolve up the RGB
from the HB level to the RGB tip. Therefore, we would expect stars of a
similar mass range, approximately 0.001 M\sol\ to occupy both these
evolutionary phases.  Using this approach we found that the HB stars
contributed about 10\% of the total cluster luminosity despite contributing
less than 0.1\% of the mass and that the stars in bin 15 of Table 4
contributed 55 - 65\% of the luminosity and slightly over 1\% of the mass.
Since post main sequence evolutionary timescales and mass luminosity
relationships are somewhat uncertain, the large contribution of these stars
to the cluster luminosity presents a problem in the calculating of synthetic
cluster M/L's. We will return to this problem later.

We chose a power-law mass function with a flattening at the faint end:
$$\phi(m) = m^{-(x+1)} ~dm~~~~~~~m \ge 0.3 \hbox{M}_\odot,\eqno(\equat)$$
$$\phi(m) = m ~dm~~~~~~~~~~~~~ m < 0.3 \hbox{M}_\odot.\eqno(\equat),$$
consistent with Pryor \et 1989, and 1991. This is similar to what has been
seen in the galactic disk (Miller \& Scalo 1979) but is opposite to what has
been derived from the luminosity functions of three globular clusters, M13,
M71, and NGC 6397, which exhibit a steepening at the low-mass end (below 0.4
M\sol) (Richer \et 1990). However, these luminosity functions were measured
at large projected radii where one would expect an enhanced number of
low-mass stars resulting from cluster dynamical evolution (i.e., mass
segregation). Cluster environmental effects (i.e. Galactic tidal fields,
bulge and disk shocking) also tend to have more drastic effects on the outer
cluster evolution, although in this case one would expect a reduction in
low-mass stars. It is fair to say that it is dangerous to make conclusions
regarding global mass functions based on photometry of stars at large
projected radii. We will show that it is not possible to reconcile the
population and KM dynamical M/L's while using a mass function that has a
steepening at the low-mass end.

Remnants were treated in the following manner: stars with initial masses of
0.87 - 1.5 M\sol, 1.65 - 4.0 M\sol\ and 4.0 - 8.0 M\sol\ become white dwarfs
with masses of 0.5 M\sol, 0.7 M\sol\ and 1.2 M\sol, respectively (Pryor \et
1991). These objects are added to the corresponding mass bins. More massive
stars, which have presumably evolved into neutron stars or black holes, are
assumed to be ejected from the cluster. This is in agreement with the large
velocities of many times the cluster escape velocity observed for these
objects in the field.  There is gathering evidence from the many
observations of millisecond pulsars (particularly in 47 Tuc, Manchester \et
1991), however, that globular clusters may be more adept at holding on to
their neutron stars than previously thought. A possible alternative,
accretion-induced collapse, a method of manufacturing neutron stars through
accretion on to a white dwarf, has also been suggested (Bailyn \& Grindlay
1990). Favoring this latter scenario is the difficulty in reconciling the
large numbers of millisecond pulsars with the small numbers of observed low
mass X-ray binaries, the more traditional projenitors of millisecond
pulsars. Clearly, these considerations would greatly complicate dynamical
models, particularly for flat mass functions for which the remnant mass
classes become more significant. As we will show, the addition of a
significant number of heavy remnants, like a steepening of the low-mass end
of the mass function, causes a deterioration in the agreement between the
dynamical and population M/L's.

\subsect{Radial Velocities}

The mass of a KM model is given by $$M = {9r_sv_s^2 \over G}
\int{{\rho \over \rho_0}r^2dr}\eqno(\equat)$$ Illingworth (1976) where $v_s$
is the scale velocity. The run of $\sigma^2_{r,i}(r)$ and $\sigma^2_{t,i}(r)$
are determined from $$\sigma_{(r,t),i}^2(r) = {\int_{|\sigma_i|
\le W(r)} f_i(\sigma_i,W)\sigma_k^2d^3\vec \sigma_i \over \int_{|\sigma_i|
\le W(r)} f_i(\sigma_i,W)d^3\vec\sigma_i},\eqno(\equat)$$ where, $\sigma_k =
\sigma_i$cos$\theta$ or $\sigma_i$sin$\theta$ for $\sigma_{r,i}$ or
$\sigma_{t,i}$, respectively, and the $i$ subscript refers to the $i^{th}$
mass class. Comparisons were made between the observed velocities and scaled
model velocity dispersions projected along the line of sight,
$$\sigma_{p,i}^2(R) = {2 \over \mu_i (R)} \int^\infty_ R{\rho_{i}(r)[(r^2-
R^2)\sigma_{r,i}^2(r)+ R^2\sigma_{t,i} ^2(r)]dr \over r(r^2-
R^2)^{1/2}},\eqno(\equat)$$ (Binney and Tremaine 1987, p. 208), where
$\mu_i$ is the surface density of the $i^{th}$ mass class. The optimal
velocity scaling and mean velocity were derived using the maximum likelihood
technique outlined in Gunn and Griffin (1979).

A serious problem in mass determinations is contamination from binary and
non-member stars, both of which tend to increase the mass estimate.  While
non-members are a virtual non-issue due to the high systemic radial velocity
of NGC 362, binaries can be especially problematic with only about one
quarter of the stars having multiple radial velocity measurements. We
attempted to deal with this problem using the technique discussed in Fischer
\et (1993).  First, using the entire data set, excepting the known
variables, the optimal $v_s$ and $v_{ave}$ are determined using equation 7
for all the KM models described below. For every star the parameter
$$\delta_i = \sqrt{{(v_{ri}-v_{ave})^2
\over {v_s^2
\sigma_p^2(R_i) + v_{err\ i}^2}}}
\eqno(\equat)$$ was tabulated. The largest $\delta_i$ obtained was
approximately 2.80.  In order to determine the likelihood of having such a
value, we performed Monte Carlo simulations of the radial velocity data. We
started with the known projected radii ($R_i$) of the program stars. The
true radius is in the range \Rc2 $\le$ r $\le$ r$_{t}$, where r$_{t}$ is the
tidal radius. If $x$ is the displacement from the mean cluster position
along the line-of-sight such that r = $\sqrt{R^2+x^2}$ then the probability
that the star is at $x$ is $$p(x) \sim
\rho_K(\sqrt{R^2+x^2}).\eqno(\equat)$$ A three-dimensional position along
with corresponding model-dependent radial and tangential velocities were
drawn at random from their respective probability distributions. The
velocity component along the line-of-sight was then determined, and an error
term, drawn from a Gaussian distribution with standard deviation equal to
the velocity error, as tabulated in Table 2, was added.  This process was
repeated, producing 1000 sets of data, each with a given mass and $r_a$ and
the same projected positions and velocity measurement errors as the original
data set. Finally the maximum likelihood technique was applied to each of
the artificial data sets and the maximum $\delta_i$ were recorded. We found
that a value of $\delta_{max} \ge 2.80$ occurred approximately 70\% of the
time implying that there is probably no high residual velocity contamination.

\subsect{Results}

The fitted KM parameters for 3 $r_s \le r_a \le \infty$ and $0.0
\le x \le 2.0$ are shown in Table 5. Column 1 is the anisotropy radius,
column 2 is the mass function slope (the first four rows are single-mass
models), column 3 is the reduced central potential, column 4 is the scale
radius, column 5 is the ratio of the tidal radius to the scale radius and
columns 6 and 7 are the reduced chi-squared for the fit ($\nu = 22$ degrees
of freedom) and the probability of exceeding this value, respectively. These
probabilities are based on 1000 Monte Carlo simulations for each parameter
set, each using a surface profile generated from the best fit model with
errors drawn from the uncertainties shown in Table 1. Hence, the
probabilities are somewhat dependent on the accuracy of the photometric
uncertainties and for this reason it is safer to view them in the relative
sense.

The values of $v_s$ are shown in column 8 of Table 5. A goodness-of-fit
statistic $$\zeta^2 = \sum{{(v_{r,k} - \vave)^2} \over (v_s^2 \sigma_{p,k}^2
+ v_{err,k}^2)}\eqno(\equat)$$ was generated for each model and is shown in
column 9. The distribution of this statistic can be extracted from the
radial velocity Monte Carlo simulations described above. We find that the
$\zeta^2$ are distributed around N, the number of radial velocity
measurements. Values of $\zeta^2$ less than N tend to indicate that the
model is too flat as a function of projected radius for the radial velocity
data, while high values imply too steep a model. The width of the $\zeta^2$
distribution appears to be dependent on the anisotropy radius but this is
probably due to the higher values of $v_s$ for the anisotropic models. These
effects are demonstrated in Fig 4. The solid histogram in the upper panel is
the distribution of $\zeta^2$ when one fits a model with $r_a = 3 r_s$ to
data with an isotropic distribution function while the dashed histogram has
$r_a = 3 r_s$ for both model and data.  Similarly, the solid histogram in the
lower panel represents an isotropic model and data with $r_a = 3 r_s$ while
the dashed histogram has isotropic orbits and data. Column 10 of Table 5
shows the probability of exceeding the observed $|\zeta^2-N|$ assuming that
the cluster velocities are specified by the model parameters indicated and
have the uncertainties tabulated in Table 2.  The greater this probability
the higher the likelihood that the cluster velocities are drawn from the
specified distribution.

Derived parameters for the KM model are displayed in Table 6: columns 1 and
2 specify the anisotropy radius and mass function slope, while columns 3 and
4 contain the central luminosity density and total cluster luminosity,
respectively. Columns 5 and 6 are the central mass density and total cluster
mass, respectively.  Columns 7 and 8 are the central and global $population$
M/L's given by $${M \over L_V} = {\int^{m_u}_{m_{l}}m\phi(m)dm \over
\int^{m_u}_{m_{l}}l(m)\phi(m)dm},\eqno(\equat)$$ where $l(m)$ is the
luminosity of a star of mass $m$ given by the Bergbusch \& VandenBerg
stellar models described above, and $m_l$ and $m_u$ are the lower and upper
mass cut-offs, respectively.

The corresponding dynamical M/L's are in columns 9 and 10, respectively.
Monte-Carlo simulations of the radial velocity data, as described above,
were use to determine the uncertainties implicit in the maximum likelihood
technique and to search for any possible systematic effects. We noticed that
the maximum likelihood method resulted in scale velocities that were biased
systematically too low but only by a few percent so they have not been
corrected. This bias tends to be larger when the number of radial velocities
is smaller.

First a comment about the single-mass models which assume that stars of all
masses have the same radial distribution. We obtain values of c and $r_s$
consistent with both Illingworth \& Illingworth (1976) and Tucholke (1992)
(both used only isotropic models) indicating that our luminosity profile is
very similar to their starcounts.  Assuming isotropy, we find a higher
central velocity dispersion (although consistent within uncertainties) than
the value of 7.8 \kms\ which was obtained by Illingworth (1976) from
integrated spectroscopy of the cluster center (this value was corrected from
the observed velocity dispersion of 7.5 \kms\ assuming an isotropic model).
Correspondingly, our isotropic single-mass mass estimate is about 30\%
higher which puts us right at the upper limit of Illingworth's 1$\sigma$
uncertainty. Our luminosity estimate is 17\% lower, resulting in a M/L =
$1.40 \pm 0.2$ M\sol/L$_V$\sol which is 55\% higher than his, a difference
significant at the 1.4$\sigma$ level.  The cluster has a mean systemic
velocity, independent of model parameters, of $223.5 \pm 2.0$ \kms where the
uncertainty reflects the accuracy of our zero-point and includes the actual
velocity errors in the standard system itself. This is in excellent
agreement with the integrated radial velocity of Dubath \et (1993) who
obtained $223.20 \pm 0.2$ \kms, and may indicate that our velocity
zero-point is more secure than 2.0 \kms.  Fig. 5 shows the surface
brightness data from Table 1 along with the isotropic and $r_a = 3 r_s$
single-mass models. Also shown as a dotted line is a typical stellar profile
having FWHM = 1.80\sec which is about 14\% of the cluster core radius.
Therefore, the core is easily resolved and the seeing will have negligible
impact on its measurement.  Finally, we have plotted a power-law with slope
equal to --1 as a long-dashed line. NGC 362 has been classified as a
possible post core-collapse cluster (PCC? in their designation scheme) by
Chernoff \& Djorgovski (1989). The criteria that these authors used to
determine if a cluster is PCC are: a small core, and ``... an unambiguous,
extended, slope $\approx -1$ power-law section of the surface brightness
profile near the center...''. Neither of these criteria are applicable to
our data and we conclude that NGC 362 is {\it not} a post core-collapse
cluster.  Unfortunately Chernoff \& Djorgovski did not include a plot of
their SBP of NGC 362.

For the multi-mass models, the trends in the goodness-of-fit to the {\it
SBP} with $r_a$ and $x$ can be categorized as follows: 1) High quality fits
can be found for the whole $x$ and most of the $r_a$ range shown in Tables 5
and 6, but not for all combinations of the parameters. 2) For low values of
$x$ the best fits occur for moderate to high amounts of anisotropy. 3) For
high values of $x$ only the isotropic models provide good fits. This is
easily explained; as the mass function steepens, one adds more low-mass
stars to the cluster, which, because of equipartition of energy migrate to
large radius giving the cluster a more extended appearance. This mimics the
effect of decreasing $r_a$ for shallower mass functions which also gives the
cluster a more extended appearance due to radial orbits traveling to larger
cluster-centric radii than equally energetic circular orbits. Overall, the
SBP is more consistent with shallow mass functions and moderately low levels
of anisotropy for the range of parameter space sampled here.

The trends are much simpler for the radial velocity modeling in that highly
anisotropic orbits are always favored, regardless of mass function.  In
fact, it is possible to rule out all but the $x=0.0$ isotropic models at a
very high level of confidence. The best overall fits to both the kinematics
and the SBP occur for $x = 0.0 - 0.5$ and $r_a = 5-10$ It should be pointed
out, however, that while the $\chi^2$ result from the fitting of the SBP,
the $\zeta^2$ arise from the fitting of radial velocities to velocity models
whose shape is fixed by the SBP leaving only a velocity scaling and mean
velocity to be determined. In other words, we have seen that the radial
velocities are most consistent with anisotropic distribution functions which
have radial velocity dispersion profiles which fall off more rapidly at
small radii than their isotropic counterparts. A similar effect can be
accomplished by decreasing the value of $r_s$ from the one most consistent
with the SBP.  Therefore, what we are really showing here is not that the
radial velocity data are only consistent with anisotropic models, but that
the only way to get consistency for all the data is to invoke moderately
anisotropic models and shallow mass functions.

One apparent problem with this set of KM models is that the central and
global population M/L's are all substantially higher than the dynamic M/L's,
and the situation is worse for the best-fitting models with $x=0.0 - 0.5$
and $r_a = 5 - 10$. These models have both global and central population
M/L's which are 1.5 - 2.5 times larger than the dynamic estimates.  How much
of a problem is this? The population M/L's are very sensitive to how the RGB
and HB are populated and different assumptions can lead to very different
results as can be seen in Pryor \et (1991) who obtain global M/L's that are
25 - 50\% lower than our tabulated values. Such a reduction would provide
significantly better agreement between the population and dynamical M/L's,
especially for $x = 0.5$. Therefore, we have arbitrarily increased the number
of stars in bin 15 by a factor of two. This is equivalent to doubling the
(uncertain) lifetimes on the RGB and the HB. Aside from the uncertainty in
these lifetimes, another justification is that the population models do not
include all the possible evolved stars (i.e. asymptotic branch stars) or the
possible existence of blue stragglers many of which may be hiding in the
core as is the case for 47 Tuc (Parasce \et 1991). The results are shown in
Tables 7 and 8 which have the same format as Tables 5 and 6, respectively.
There is very little change in most of the derived and fitted parameters or
the quality of the fits for a given set of parameters.  However, one can see
that the desired effect has been achieved, the values of the global and
central population M/L's are reduced and the agreement, particularly for $x
= 0.5 - 1.0$, is considerably improved.  Clearly, with similar manipulations
one could achieve good agreement with virtually any $x$ and this tends to
make {\it comparisons between dynamic and population M/L's a fairly weak
constraint}.  The situation could be greatly improved with complete number
counts of the evolved stars; a difficult task requiring high resolution
imaging.

To summarize, the best fits to both the kinematic data and surface
photometry occur for a flat mass function with moderate amounts of
anisotropy. The cluster luminosity is around 1.70 $\pm 0.1 \times 10^5$
L$_V$\sol\ and the mass ranges from approximately $2.5 - 3.5 \times 10^5$
M\sol\ for a global mass-to-light in the range M/L$_V$ = 1.5 - 2.0
M\sol/L$_V$\sol with a mass function near $x = 0.5$.  Unlike the findings of
Richer \et (1990) for 3 other clusters, our data is inconsistent with a sharp
upturn in the mass function at the low-mass end. Such an upturn would result
in a large increase in the population M/L requiring a large increase in the
number of giants or a decrease in the number of stellar remnants both of
which would be hard to justify. It would also be possible to get good
agreement between the SBP and the kinematics for mass functions less than
zero and isotropic orbits (not shown in tables) However, this results in a
large divergence between population and dynamic M/L's.  Including large
numbers of heavy remnants (black holes, neutron stars) results in a similar
problem.

Recently Capaccioli \et (1991) have shown that there exists a fairly strong
correlation between the slope of cluster mass functions and both the
clusters' distance from the Galactic center (R$_{GC}$) and above the Galactic
disk (Z).  There are several reasons why one would expect shallower cluster
mass function near the Galactic disk and center. Perhaps the most compelling
is heating of the cluster through disk and tidal shocking, although Galactic
tidal fields will also play a role. The effect is to preferentially
evaporate low-mass stars, which tend to be more loosely bound to the cluster,
resulting in a present day mass function (PDMF) which is flatter than the
initial mass function (IMF). NGC 362 with (m--M)$_0$ = 14.77 mag, $l =$
301.53\deg and $b=$ -46.25\deg, has R$_{GC} = 9.6$ kpc and Z = 6.5 kpc
(assumes the distance from the sun to the galactic center is 8 kpc).
Therefore, NGC 362 has a similar value of $x$ and R$_{GC}$ to M92 and hence
agrees fairly well with the weaker $x$ - R$_{GC}$ correlation.  However, it
tends to be an outlier in the stronger $x$ - Z correlation (assuming it was
extrapolated to 6.5 kpc) requiring a value of about $x = 1.0$ for
consistency. NGC 362 has a larger value of Z than all but one cluster in the
Capaccioli \et compilation; the largest is M3 with Z = 10.2 which also lies
below the correlation. The evidence from these two high Z clusters may
indicate that the relationship is flattening.  Perhaps this implies that NGC
362 is sufficiently far from the disk that tidal shocking is relatively
unimportant to its dynamical evolution and that its mass function has
evolved to a lesser degree than those of the low Z clusters. A more
interesting correlation to search for, in terms of disk and bulge shocking,
would be the one between $x$ and various cluster orbital parameters,
particularly the clusters' perigalacticon. The mass functions should be more
sensitive to these parameters, as they more truly reflect the
environmentally driven aspects of cluster evolution (i.e. see Aguilar \et
1988).

\subsect{Relaxation Timescales}

As mentioned above, a real cluster cannot achieve complete equipartition of
energy regardless of its age. However, since, through the use of KM models,
we are assuming a large degree of mass segregation, it is necessary to
justify this in terms of the cluster's evolutionary state. Two relevant
timescales are the oft-quoted central relaxation time

$$t_{r\circ} = (1.55 \times 10^7
\hbox{yr})\left({r_s
\over \hbox{pc}}\right)^2\left({v_s \over \hbox{km s}^{-1}}\right)
\left({M_\odot
\over \left< m \right>} \right)\left[\log(0.5 M / \left< m
\right>)\right]^{-1} = (0.05 - 1.0) \times 10^8 \hbox{yr}\eqno(\equat),$$
(Lightman and Shapiro 1978) and the half mass relaxation time $$t_{rh}=(8.92
\times 10^8 \hbox{yr})\left({M
\over 10^6 M_\odot}\right)^{1/2}\left({r_h \over \hbox{pc}
}\right)^{3/2}\left({M_\odot \over
\left< m \right>}\right)\left[\log(0.4M/\left< m \right>)\right]^{-1} = (0.7
- 2.3) \times 10^9 \hbox{yr}\eqno(\equat)$$ (Spitzer and Hart 1971).

The ranges in relaxation time correspond to the parameter sets described
above. It appears that the system is largely relaxed therefore necessitating
the use of multi-mass models.

\sect{Conclusions}

In this paper we have examined the internal dynamics of the globular
cluster NGC 362 using V band CCD images and echelle spectra of 208 member
red giants.

\itemn The new oxygen-enhanced isochrones of Bergbusch \& VandenBerg (1992)
with [Fe/H] = -1.03 were applied to the fiducial main sequence and sub-giant
branch (Bolte 1987) of NGC 362 yielding an age of 13 Gyr, an absolute
distance modulus (m--M)$_0$ = 14.77 mag, and reddening E(B-V) = 0.0 mag.

\itemn A CCD surface brightness profile (SBP) extending out to approximately
20 pc was constructed from a mosaic of four V band images. After determining
that NGC 362 is not post core-collapse, single- and multi-mass King Michie
models were applied to the data for a range of mass function slopes $0.0 \le
x \le 2.0$ and anisotropy radii $3.0 \le r_a (r_s)
\le \infty$.  The quality of the fits varied as a function of both these
parameters; for small values of $x$, anisotropic models were favored while
for steeper mass functions isotropic orbits provided superior fits. The
derived total cluster luminosity is 1.70 $\pm 0.1 \times 10^5$ L$_V$\sol.

\itemn A total of 285 stellar spectra of 215 stars were obtained; 50 were
observed twice, one observed three time, one observed ten times, and one
observed eleven times. Three stars were obvious non-members and four stars
had strong evidence for radial velocity variability which we conclude is
likely due to the presence of binary stars. The implied cluster binary
fraction is 0.15 for circular orbits or 0.27 for an orbital distribution
function $f(e) = e$. This is a considerably higher binary fraction than
previous cluster studies have found. While it may indicate that NGC 362 has
a high binary fraction, it might also be a result of the way we selected
candidates for repeat measurements. Furthermore, this fraction is based
on only two measurements of each of the candidate binaries.

\itemn The 208 remaining stars with radial velocity measurements were
incompatible with King-Michie models having isotropic velocity dispersion
and luminosity profiles consistent with the SBP. Therefore, the best
agreement with both the kinematic data and the SBP were for shallow mass
functions $x = 0.0 - 0.5$ and intermediate amounts of velocity anisotropy.
In this best-fit range, the derived cluster mass was M = $2.5 - 3.5 \times
10^5$ M\sol\ for a global mass-to-light ratio of M/L$_V$ = 1.5 - 2.0
M\sol/L$_V$\sol.

\itemn The values of $x$ and Z, the cluster height above the Galactic disk,
do not lie on the correlation (assuming it was extrapolated to Z = 6.5 kpc)
between these two parameters found by Capaccioli \et (1991). NGC 362 has a
mass function which is too flat, possibly implying that the cluster
evolutionary drivers (i.e. disk and bulge shocking, tidal fields) play a
relatively small role at its location in the Galaxy.

\itemn Our value of $x$ differs from the sharp turn-ups in mass
functions derived from some deep luminosity function work reported in Richer
\et (1990) for the clusters M13, M71, and NGC 6397.

\itemn In the future we will attempt non-parametric modeling of the cluster
radial velocities as described by Merritt (1992).

\centerline{ACKNOWLEDGMENTS}

P.F. would like to acknowledge the Natural Sciences and Engineering Research
Council (NSERC) and AT\&T for post-doctoral fellowships. This work was
undertaken while D.L.W. was a NSERC University Research Fellow. Partial
support for this work was provided by NASA through grant \# HF-1007.01-90A
awarded by the Space Telescope Science Institute which is operated by the
Association of Universities for Research in Astronomy, Inc., for NASA under
contract NAS5-26555. We would like to thank Ben Dorman for some helpful
discussions regarding horizontal branches and P.F. would like to thank
Robert Lupton for his generous gift of a free copy of sm.

\vfill\eject

{\everypar={\parindent=0pt\parskip=0pt\hangindent=20pt\hangafter=1}

\noindent
\ref{Aguilar, L., Hut, P., \& Ostriker, J. P. 1988}{\apj}{335}{720}

\ref{Alcaino, G. 1976}{\aas}{26}{359}

\ref{Alcaino, G., Liller, W., \& Hazen, M.L. 1988}{\aj}{96}{139}

\ref{Bailyn, C. D., \& Grindlay, J. E. 1990}{\apj}{353}{159}

\ref{Bergbusch, P. \& Vandenberg, D. A., 1992}{\apjs}{81}{161}

\fullref{Binney J., \& Tremaine, S. 1987, in {\it Galactic Dynamics},
(Princeton: Princeton University Press)}

\ref{Bolte, M. 1987}{\apj}{315}{469}

\ref{Burstein D., \& Heiles, C. 1978}{\apj}{225}{40}

\ref{Chernoff, D. F., \& Djorgovski, S. 1989}{\apj}{339}{904}

\ref{C\^ot\'e, P., Welch, D. L., Mateo, M., Fischer, P., and Madore, B. F.
1991}{\aj}{101}{1681}

\fullref{C\^ot\'e, P., Welch, D. L., Fischer, P., and Irwin, M. J. 1993}

\ref{Da Costa, G. S., Freeman, K. C. 1976}{\apj}{206}{128}

\ref{Capaccioli, M., Ortolani, S., \& Piotto, G. 1991}{\aa}{244}{298}

\ref{Dickens, R. J., Croke, B. F. W., Cannon, R. D., \& Bell, R. A.
1991}{Nature}{351}{212}

\fullref{Djorgovski, S. 1988, in The Harlow-Shapley Symposium on
Globular Cluster Systems in Galaxies (IAU Symposium No. 126)}, eds. J. E.
Grindlay and A. G. Davis Philip, (Dordrecht: Reidel), p. 333

\ref{Dorman, B, 1992}{\apjs}{81}{221}

\fullref{Dubath, P., Meylan, G., \& Mayor,  M. 1993 \aa Submitted}

\ref{Fischer, P, Welch, D. L., \& Mateo, M. 1993}{\aj}{105}{938}

\ref{Fischer, P, Welch, D. L., C\^ot\'e, P., Mateo, M., \& Madore, B. F.
1992}{\aj}{103}{857}

\ref{Gunn, J. E. \& Griffin, R. F. 1979}{\aj}{84}{752}

\ref{Harris, W.E. 1982}{\apjs}{50}{573}

\ref{Heggie, D. C. 1975}{\mnras}{173}{729}

\ref{Hut, P., \et 1992}{\pasp}{104}{981}

\ref{Illingworth, G. 1976}{\apj}{204}{73}

\ref{Illingworth, G., \& Illingworth, W. 1976}{\apjs}{30}{227}

\ref{Inagaki, S. \& Saslaw, W. C. 1985}{\apj}{292}{339}

\ref{Jenkner, H., Lasker, B. M., Sturch, C. R., McLean, B. J., Shara, M. M.,
\& Russell, J. L. 1990}{\aj}{99}{2019}

\ref{King, I. R. 1966}{\aj}{71}{64}

\ref{Lasker, B. M., Sturch, C. R., McLean, B. J., Russell, J. L., Jenkner,
H., \& Shara, M. M. 1990}{\aj}{99}{2019}

\ref{Lightman, A. P. \& Shapiro, S. L. 1978}{Rev. Mod. Phys.}{50}{437}

\ref{Lupton, R. H., Gunn, J. E. \& Griffin, R. F. 1987}{\aj}{93}{1114}

\ref{Mayor, M., Imbert, M., Andersen, J., Ardeberg, A., Benz, W., Lindgren,
H., Martin, N., Maurice, E., N\"ordstrom, B., and Pr\'evot, L.
1984}{\aa}{134}{118}

\fullref{Merritt, D. 1992 in Dynamics of Globular Clusters, eds. S.
Djorgovski and G. Meylan, in press}

\ref{Meylan, G. \& Mayor, M. 1986}{\aa}{166}{122}

\ref{Meylan, G. \& Mayor, M. 1991}{\aa}{250}{113}

\ref{Michie, R. W. 1963}{\mnras}{126}{499}

\fullref{Mihalas, D., \& Binney, J. 1981, in {\it Galactic Astronomy Structure
and Kinematics}, (W. H. Freeman and Company: New York)}

\ref{Miller, G. E. \& Scalo, J. M. 1979}{\apjs}{41}{513}

\ref{Paresce F. \et 1991}{Nature}{352}{297}

\ref{Peterson, R. C., Seitzer, P., \& Cudworth, K. M. 1989}{\apj}{347}{251}

\ref{Pryor, C., McClure, R. D., Fletcher, J. M., Hartwick, F. D. A., \&
Kormendy, J. 1986}{\aj}{91}{546}

\ref{Pryor, C. P., Latham, D. W., and Hazen, M. L. 1988}{\aj}{96}{123}

\ref{Pryor, C., McClure, R. D., Fletcher, J. M., \& Hesser, J. E. 1989}{\aj}
{98}{596}

\ref{Pryor, C., McClure, R. D., Fletcher, J. M., \& Hesser, J. E. 1991}{\aj}
{102}{1026}

\ref{Richer, H. B., Fahlman, G. G., Buonanno, R., \& Fusi Pecci, F.
1990}{\apj}{359}{L11}

\ref{Russell, J. L., Lasker, B. M., McLean, B. J., Sturch, C. R., \&
Jenkner, H. 1990}{\aj}{99}{2059}

\ref{Sawyer-Hogg, H. 1973}{Publ. D.D.O.}{3}{6}

\ref{Shawl, S.J. \& White, R.E.}{1986}{\aj}{91}{312}

\ref{Spitzer, L. \& Hart, M. H. 1971}{\apj}{164}{399}

\ref{Suntzeff, N.B., Friel, E., Klemola, A., Kraft, R.P., and Graham, J.A.
1986}{\aj}{91}{275}

\fullref{Tody, D. 1986 in {\it The IRAF Data Reduction and Analysis System,
Instrumentation in Astronomy VI}, ed. D. L. Crawford, Proc. SPIE,
p. 713}

\ref{Tonry, J., \& Davis, M. 1979}{\aj}{84}{1511}

\ref{Tucholke, H. J. 1992}{\aa}{}{513}

\ref{VandenBerg, D. A. 1992}{\apj}{391}{685}

\ref{VandenBerg, D. A., Bell, R. A.  1992}{\apjs}{58}{561}

\ref{Welch, D. L., Mateo, M., C\^ot\'e P., Fischer, P., \& Madore, B. F.
1991}{\aj}{101}{490}

\ref{White, R.E. \& Shawl, S.J. 1987}{ApJ}{317}{246}

}

\vfill\eject

\figc Mean radial velocity vs. projected radius (upper panel) and versus
position angle (lower panel) for the 212 member stars. The solid lines are
the mean velocity, $\vave = 223.5 \pm 2.0$ km s$^{-1}$.

\figc The difference in median velocity for stars on either side of an axis
at the specified position angle. Also shown is the best fit sine function
with the parameters shown at the top of the plot.

\figc A plot of the fiducial main sequence ridge (dotted line) from Bolte
(1987). The solid lines are isochrones from Bergbusch \& VandenBerg (1992)
for [Fe/H] = --1.03, Y=0.2368, (m--M) = 14.77 mag, and E(B--V) = 0 mag with
ages of $\tau = $ 12, 13, and 14 Gyr. The long-dashed line is an isochrone
from VandenBerg
\& Bell (1985) with for [Fe/H] = -0.79, Y=0.3, (m--M) = 14.60, and E(B--V) =
0 mag with an age of $\tau =14$ Gyr.

\figc The solid histogram in the upper panel is the distribution of
$\zeta^2$ when one fits a model with $r_a = 3 r_s$ to data with an isotropic
distribution function while the dashed histogram has $r_a = 3 r_s$ for both
model and data.  Similarly, the solid histogram in lower panel represents
and isotropic model and data with $r_a = 3 r_s$ while the dashed histogram
has isotropic orbits and data.

\figc The surface brightness data (squares) along with the isotropic (solid
line) and $r_a = 3 r_s$ (dashed line) single-mass models. The dotted line is
a typical stellar profile and the long-dashed line is a power law
relationship with slop equal to --1.

\def \rlap #1{\hbox to Opt{#1\hss}}
\def\ps{\phantom{0}}
\def\pss{\phantom{00}}
\def\pd{\phantom{.}}
\baselineskip=12pt

\pageinsert
\tabh{\taone}{Surface Photometry}
\span\tcoli && \span\tcol{.3} \cr
\topper

\hfil R$_\theta$ & R & $\mu_V$ & R$_{min}$ & R$_{max}$ \cr
\hfil ($\prime$) & (pc) & L$_V$\sol\ pc$^{-2}$ & (pc) & (pc) \cr
\spacer

0.023 & ~0.06 & 32127.0 $\pm$1535.0 &~0.00 &~0.08 \cr
0.037 &  ~0.10 & 32560.0 $\pm$4193.0 &~0.08 &~0.11 \cr
0.046 &  ~0.12 & 33119.0 $\pm$5299.0 &~0.11 &~0.13 \cr
0.058 &  ~0.15 & 33167.0 $\pm$6279.0 &~0.13 &~0.17 \cr
0.073 &  ~0.19 & 32920.0 $\pm$6659.0 &~0.17 &~0.21 \cr
0.092 &  ~0.24 & 32034.0 $\pm$9660.0 &~0.21 &~0.27 \cr
0.115 &  ~0.30 & 22848.0 $\pm$6562.0 &~0.27 &~0.33 \cr
0.145 &  ~0.38 & 21508.0 $\pm$3666.0 &~0.33 &~0.42 \cr
0.183 &  ~0.48 & 14495.0 $\pm$2926.0 &~0.42 &~0.53 \cr
0.230 &  ~0.60 & 15507.0 $\pm$1261.0 &~0.53 &~0.67 \cr
0.290 &  ~0.76 & 14659.0 $\pm$2308.0 &~0.67 &~0.84 \cr
0.364 &  ~0.95 & ~9624.0 $\pm$1062.0 &~0.84 &~1.06 \cr
0.459 &  ~1.20 & ~5206.0 $\pm$~311.0 &~1.06 &~1.33 \cr
0.578 &  ~1.51 & ~3634.0 $\pm$~282.0 &~1.33 &~1.68 \cr
0.727 &  ~1.90 & ~2581.0 $\pm$~258.0 &~1.68 &~2.11 \cr
0.916 &  ~2.40 & ~1501.0 $\pm$~104.0 &~2.11 &~2.66 \cr
1.153 &  ~3.02 & ~~983.0 $\pm$~~69.0 &~2.66 &~3.35 \cr
1.452 &  ~3.80 & ~~565.0 $\pm$~~51.0 &~3.35 &~4.22 \cr
1.828 &  ~4.78 & ~~362.0 $\pm$~~58.0 &~4.22 &~5.31 \cr
2.301 &  ~6.02 & ~~173.0 $\pm$~~34.0 &~5.31 &~6.68 \cr
2.897 &  ~7.58 & ~~105.0 $\pm$~~15.0 &~6.68 &~8.41 \cr
3.647 &  ~9.54 & ~~~56.0 $\pm$~~~7.8 &~8.41 &10.59 \cr
4.592 &  12.01 & ~~~29.7 $\pm$~~~6.6 &10.59 &13.33 \cr
5.780 &  15.12 & ~~~11.2 $\pm$~~~9.4 &13.33 &16.78 \cr
7.270 &  19.02 & ~~~ 3.2 $\pm$~~~5.3 &16.78 &21.13 \cr
\sp
\endinsert
\vfill\dosupereject

\pageinsert
\baselineskip=10pt
\eightpoint\rm
\tabh{\taone}{Radial Velocities}
\span\tcoli && \span\tcol{0.0} \cr
\topper

Star & $\alpha$ & $\delta$ & R & PA & HJD & $v_r$ & $\overline{v_r}$ & $\chi^2$
& $\nu$ & Comments \cr
& (2000.0) & (2000.0) & (arcsec) & ($^{\circ}$) & -2400000 & (km\ s$^{-1}$) &
(\kms) \cr
\spacer

H1106&01:03:10.50&-70:49:10.8&104.78&349.87&47872.6208&223.89~$\pm$~4.70&&&&\cr
H1111&01:03:07.42&-70:49:21.3&~98.55&340.06&47872.6253&222.53~$\pm$~1.72&&&&\cr
H1114&01:03:03.80&-70:48:45.6&138.27&338.15&47872.6232&223.97~$\pm$~2.50&&&&\cr
H1117&01:03:04.05&-70:49:36.0&~92.70&327.21&47872.6271&233.71~$\pm$~1.42&&&&\cr
&&&&&48605.6503&237.82~$\pm$~1.78&235.31~$\pm$~1.11&~~3.26&~1&\cr
H1137&01:02:58.50&-70:49:44.2&104.26&311.96&47872.6287&224.70~$\pm$~0.63&&&&\cr
H1154&01:02:52.03&-70:50:10.7&117.57&291.54&47872.6303&224.82~$\pm$~1.19&&&&\cr
H1157&01:02:55.94&-70:50:17.4&~97.21&292.05&47872.5813&218.33~$\pm$~0.64&&&&\cr
H1159&01:02:53.14&-70:50:30.6&106.44&282.63&47872.5830&217.20~$\pm$~0.59&&&&\cr
H1165&01:02:53.11&-70:50:49.5&104.08&272.41&47872.5851&220.92~$\pm$~1.46&&&&\cr
H1166&01:02:51.41&-70:50:50.7&112.40&271.61&47872.5868&223.81~$\pm$~1.07&&&&\cr
H1168&01:02:45.18&-70:50:57.0&143.04&268.72&47872.5893&223.23~$\pm$~2.89&&&&\cr
H1204&01:02:46.62&-70:51:16.7&137.79&260.44&47873.5796&219.69~$\pm$~9.58&&&&RR
Lyr (V11)\cr
H1206&01:02:54.49&-70:51:23.4&101.54&253.10&47873.5715&219.79~$\pm$~1.78&&&&\cr
H1207&01:02:53.65&-70:51:28.0&106.87&251.38&47873.5740&224.17~$\pm$~2.78&&&&\cr
H1211&01:02:50.10&-70:51:46.1&129.70&246.24&47873.5764&218.04~$\pm$~1.51&&&&\cr
H1216&01:02:58.56&-70:51:53.1&~97.20&232.49&47873.6166&227.32~$\pm$~0.83&&&&\cr
H1218&01:03:01.50&-70:52:04.9&~94.66&221.43&47873.6131&216.64~$\pm$~1.32&&&&\cr
H1225&01:03:02.11&-70:52:31.6&114.43&211.40&47873.6149&231.07~$\pm$~1.17&&&&\cr
H1238&01:03:10.02&-70:52:56.3&124.10&189.62&47873.6107&231.89~$\pm$~2.77&&&&\cr
&&&&&48607.5549&228.34~$\pm$~2.55&229.97~$\pm$~1.88&~~0.89&~1&\cr
H1240&01:03:11.99&-70:52:52.0&118.57&185.35&47873.6085&221.12~$\pm$~1.83&&&&\cr
%% FOLLOWING LINE CANNOT BE BROKEN BEFORE 80 CHAR
H1243&01:03:13.04&-70:53:06.8&132.98&182.54&47873.6070&~~4.02~$\pm$~0.60&&&&Non-member\cr
H1244&01:03:13.64&-70:53:19.3&145.38&181.16&47873.6053&231.02~$\pm$~3.77&&&&\cr
&&&&&48607.5596&232.42~$\pm$~1.84&232.15~$\pm$~1.65&~~0.11&~1&\cr
H1309&01:03:18.33&-70:53:16.3&143.76&171.97&47874.6030&227.63~$\pm$~1.64&&&&\cr
H1325&01:03:32.74&-70:52:43.2&142.17&140.25&47874.5928&220.47~$\pm$~2.31&&&&\cr
H1330&01:03:30.96&-70:52:06.4&109.60&131.41&47874.5582&220.92~$\pm$~1.13&&&&\cr
H1333&01:03:34.00&-70:52:04.4&120.04&125.98&47874.5599&217.72~$\pm$~1.30&&&&\cr
H1334&01:03:38.37&-70:52:04.5&138.07&120.78&47874.5895&214.99~$\pm$~0.81&&&&\cr
&&&&&48605.6711&215.19~$\pm$~1.08&215.06~$\pm$~0.65&~~0.02&~1&\cr
H1340&01:03:33.12&-70:51:37.8&102.71&115.31&47874.5616&216.60~$\pm$~1.50&&&&\cr
H1341&01:03:31.69&-70:51:31.6&~93.74&113.71&47874.5633&224.46~$\pm$~1.39&&&&\cr
H1342&01:03:35.52&-70:51:28.6&110.28&108.35&47874.5798&224.48~$\pm$~1.12&&&&\cr
H1344&01:03:36.72&-70:51:20.6&113.77&103.59&47874.5780&223.49~$\pm$~1.04&&&&\cr
H1348&01:03:42.06&-70:51:25.2&140.40&102.92&47874.5819&213.25~$\pm$~2.15&&&&\cr
&&&&&48605.6148&219.41~$\pm$~1.26&217.84~$\pm$~1.09&~~6.11&~1&Probable
binary\cr
H1351&01:03:42.75&-70:51:06.5&140.85&~~95.1&47874.5841&218.42~$\pm$~1.69&&&&\cr
H1354&01:03:32.97&-70:51:03.4&~92.65&~~95.8&47874.5763&224.30~$\pm$~1.05&&&&\cr
H1409&01:03:35.57&-70:50:30.9&107.48&~~77.6&47871.5569&230.53~$\pm$~1.69&&&&\cr
H1412&01:03:34.83&-70:50:17.9&107.57&~~70.4&47871.5466&225.96~$\pm$~1.10&&&&\cr
H1415&01:03:37.20&-70:50:05.6&122.93&~~66.8&47871.5705&220.15~$\pm$~0.95&&&&\cr
H1417&01:03:33.77&-70:50:13.0&104.49&~~66.9&47871.5447&214.15~$\pm$~2.32&&&&\cr
&&&&&48605.6303&212.90~$\pm$~1.57&213.29~$\pm$~1.30&~~0.20&~1&\cr
H1419&01:03:30.60&-70:50:10.3&~91.60&~~61.5&47871.5428&235.00~$\pm$~0.86&&&&\cr
&&&&&48605.6321&243.52~$\pm$~1.84&236.53~$\pm$~0.78&~17.60&~1&Probable
binary\cr
H1422&01:03:28.79&-70:49:52.6&~94.31&~~49.4&47871.5268&231.82~$\pm$~0.86&&&&\cr
\spc
\baselineskip=12pt
\endinsert
\vfill\dosupereject

\pageinsert
\baselineskip=10pt
\eightpoint\rm
\tabh{\taonen}{Radial Velocities}
\span\tcoli && \span\tcol{0.0} \cr
\topperc

Star & $\alpha$ & $\delta$ & R & PA & HJD & $v_r$ & $\overline{v_r}$ & $\chi^2$
& $\nu$ & Comments \cr
& (2000.0) & (2000.0) & (arcsec) & ($^{\circ}$) & -2400000 & (km\ s$^{-1}$) &
(\kms) \cr
\spacer

H1423&01:03:32.96&-70:49:37.1&120.01&~~50.2&47867.5250&233.98~$\pm$~0.63&&&&\cr
&&&&&47868.5215&232.25~$\pm$~0.66&&&&\cr
&&&&&47869.5228&233.67~$\pm$~0.62&&&&\cr
&&&&&47870.5249&233.63~$\pm$~0.58&&&&\cr
&&&&&47871.5235&233.41~$\pm$~0.62&&&&\cr
&&&&&47871.5723&234.34~$\pm$~0.72&&&&\cr
&&&&&47872.5655&233.48~$\pm$~0.62&&&&\cr
&&&&&47873.5581&233.36~$\pm$~0.61&&&&\cr
&&&&&47874.5394&233.91~$\pm$~0.71&&&&\cr
&&&&&48605.6183&234.03~$\pm$~0.80&&&&\cr
&&&&&48607.5786&231.35~$\pm$~0.74&233.42~$\pm$~0.20&~14.76&10&\cr
H1435&01:03:23.07&-70:49:14.9&108.17&~~23.7&47871.5845&228.99~$\pm$~2.82&&&&\cr
H1441&01:03:21.70&-70:48:40.1&138.80&~~15.3&47871.5826&228.93~$\pm$~0.82&&&&\cr
H1448&01:03:17.75&-70:49:27.3&~88.36&~~11.2&47871.5886&215.98~$\pm$~2.31&&&&\cr
H1449&01:03:16.63&-70:49:15.3&~99.35&~~6.81&47871.5865&228.91~$\pm$~1.43&&&&\cr
H2104&01:03:07.48&-70:47:36.6&200.14&350.41&47872.6159&216.77~$\pm$~1.66&&&&\cr
&&&&&48605.6447&221.02~$\pm$~1.96&218.55~$\pm$~1.27&~~2.74&~1&\cr
H2106&01:03:03.78&-70:48:10.1&171.76&342.52&47872.6185&219.77~$\pm$~2.48&&&&\cr
H2108&01:03:00.07&-70:47:50.8&196.01&339.10&47872.6136&224.26~$\pm$~0.92&&&&\cr
H2109&01:02:58.66&-70:48:07.5&183.30&335.21&47872.6111&212.10~$\pm$~2.61&&&&\cr
&&&&&48605.6474&220.09~$\pm$~2.57&216.16~$\pm$~1.83&~~4.76&~1&\cr
%% FOLLOWING LINE CANNOT BE BROKEN BEFORE 80 CHAR
H2113&01:02:53.11&-70:48:54.6&158.35&318.87&47872.6087&~29.77~$\pm$~1.97&&&&Non-member\cr
H2115&01:02:43.30&-70:48:47.8&197.83&309.56&47872.6071&225.63~$\pm$~0.80&&&&\cr
H2117&01:02:40.56&-70:49:00.2&201.13&304.37&47872.6051&220.29~$\pm$~3.02&&&&\cr
H2122&01:02:45.25&-70:49:59.1&152.89&290.97&47872.6026&219.50~$\pm$~2.61&&&&\cr
H2124&01:02:39.09&-70:50:17.4&176.84&281.86&47872.5994&217.55~$\pm$~1.95&&&&\cr
&&&&&48605.6557&221.49~$\pm$~1.85&219.62~$\pm$~1.34&~~2.15&~1&\cr
H2127&01:02:37.62&-70:50:36.8&181.04&275.36&47872.5913&224.66~$\pm$~0.68&&&&\cr
H2205&01:02:38.04&-70:51:21.4&180.21&261.17&47873.5816&237.90~$\pm$~0.97&&&&\cr
&&&&&48605.6602&198.78~$\pm$~1.02&219.32~$\pm$~0.70&772.41&~1&Probable
binary\cr
H2206&01:02:39.07&-70:51:22.3&175.35&260.63&47873.5834&218.32~$\pm$~1.50&&&&\cr
H2212&01:02:34.06&-70:51:57.0&207.45&252.23&47873.5899&220.11~$\pm$~1.35&&&&\cr
H2213&01:02:45.10&-70:52:10.3&162.39&241.90&47873.5916&217.86~$\pm$~0.79&&&&\cr
H2220&01:02:54.36&-70:53:01.2&160.46&217.49&47873.5936&223.25~$\pm$~1.74&&&&\cr
H2221&01:03:00.94&-70:53:17.4&157.65&204.48&47873.5958&225.07~$\pm$~1.93&&&&\cr
H2222&01:03:06.28&-70:53:37.7&168.36&193.42&47873.5982&240.63~$\pm$~4.28&&&&\cr
&&&&&48605.6653&228.08~$\pm$~3.12&232.43~$\pm$~2.52&~~5.61&~1&Probable
binary\cr
H2223&01:03:09.02&-70:53:20.6&148.88&189.92&47873.6032&233.72~$\pm$~1.51&&&&\cr
&&&&&48605.6627&230.59~$\pm$~1.78&232.41~$\pm$~1.15&~~1.80&~1&\cr
H2224&01:03:09.44&-70:53:43.4&171.08&187.92&47873.6010&226.26~$\pm$~2.84&&&&\cr
H2302&01:03:18.64&-70:53:52.9&180.25&173.12&47874.6006&216.48~$\pm$~1.12&&&&\cr
&&&&&48605.6672&216.61~$\pm$~1.16&216.54~$\pm$~0.81&~~0.01&~1&\cr
H2307&01:03:24.18&-70:53:56.8&189.26&165.06&47874.5989&231.45~$\pm$~1.17&&&&\cr
&&&&&48605.6690&233.59~$\pm$~1.32&232.39~$\pm$~0.88&~~1.47&~1&\cr
H2309&01:03:29.03&-70:53:33.2&175.06&155.49&47874.5970&221.96~$\pm$~1.38&&&&\cr
H2311&01:03:27.96&-70:53:11.1&152.84&153.84&47874.5950&230.08~$\pm$~1.54&&&&\cr
H2324&01:03:43.85&-70:52:20.5&169.40&120.78&47874.5907&223.77~$\pm$~2.54&&&&\cr
H2334&01:03:55.61&-70:51:00.9&203.69&~~92.0&47874.5865&217.97~$\pm$~2.63&&&&\cr
&&&&&48607.5728&219.77~$\pm$~2.36&218.97~$\pm$~1.76&~~0.26&~1&\cr
\spc
\baselineskip=12pt
\endinsert
\vfill\dosupereject

\pageinsert
\baselineskip=10pt
\eightpoint\rm
\tabh{\taonen}{Radial Velocities}
\span\tcoli && \span\tcol{0.0} \cr
\topperc

Star & $\alpha$ & $\delta$ & R & PA & HJD & $v_r$ & $\overline{v_r}$ & $\chi^2$
& $\nu$ & Comments \cr
& (2000.0) & (2000.0) & (arcsec) & ($^{\circ}$) & -2400000 & (km\ s$^{-1}$) &
(\kms) \cr
\spacer

H2401&01:03:46.66&-70:50:48.6&159.64&~~88.1&47871.5605&233.98~$\pm$~1.11&&&&\cr
&&&&&48605.6245&234.16~$\pm$~1.08&234.07~$\pm$~0.77&~~0.01&~1&\cr
H2403&01:03:51.21&-70:50:27.7&183.85&~~81.8&47871.5624&226.46~$\pm$~1.93&&&&\cr
H2404&01:03:54.46&-70:50:22.6&200.43&~~81.0&47871.5644&235.59~$\pm$~1.56&&&&\cr
&&&&&48605.6271&235.15~$\pm$~2.26&235.45~$\pm$~1.28&~~0.03&~1&\cr
H2410&01:03:50.50&-70:49:22.4&200.65&~~62.9&47871.5666&221.23~$\pm$~2.18&&&&\cr
H2411&01:03:40.21&-70:49:41.8&146.82&~~60.6&47871.5686&223.98~$\pm$~1.65&&&&\cr
H2417&01:03:32.82&-70:48:53.1&151.59&~~37.1&47871.5742&222.15~$\pm$~1.90&&&&\cr
H2418&01:03:32.13&-70:48:40.2&160.17&~~33.4&47871.5780&218.37~$\pm$~1.53&&&&\cr
H2419&01:03:36.50&-70:48:47.6&167.29&~~40.9&47871.5761&232.23~$\pm$~1.67&&&&\cr
&&&&&48605.6354&233.06~$\pm$~1.35&232.73~$\pm$~1.05&~~0.15&~1&\cr
H2423&01:03:34.89&-70:48:07.7&194.91&~~31.5&47871.5799&223.10~$\pm$~0.92&&&&\cr
H2431&01:03:24.04&-70:47:50.4&189.80&~~14.7&47871.5813&224.74~$\pm$~0.77&&&&\cr
a    &01:03:13.64&-70:51:13.7&~19.97&188.50&47867.5287&220.56~$\pm$~0.71&&&&\cr
&&&&&47868.5365&219.70~$\pm$~0.70&220.12~$\pm$~0.50&~~0.74&~1&\cr
aa   &01:03:15.46&-70:51:06.5&~13.91&154.44&47868.5313&204.70~$\pm$~1.52&&&&\cr
ab   &01:03:15.47&-70:51:09.6&~16.78&158.86&47868.5328&216.94~$\pm$~1.04&&&&\cr
ac   &01:03:14.67&-70:51:15.0&~21.16&174.26&47868.5346&227.01~$\pm$~0.81&&&&\cr
ad   &01:03:14.22&-70:51:15.8&~21.85&180.26&47868.5383&233.12~$\pm$~0.82&&&&\cr
ae   &01:03:13.32&-70:51:22.3&~28.71&189.07&47868.5402&218.52~$\pm$~0.70&&&&\cr
af   &01:03:15.13&-70:51:26.8&~33.14&172.41&47868.5420&208.11~$\pm$~0.74&&&&\cr
&&&&&47874.5410&208.79~$\pm$~0.91&&&&\cr
&&&&&48607.5508&208.92~$\pm$~0.82&208.56~$\pm$~0.47&~~0.63&~2&\cr
ag   &01:03:15.01&-70:51:32.5&~38.74&174.39&47868.5437&223.09~$\pm$~1.04&&&&\cr
&&&&&47874.5427&222.08~$\pm$~1.27&222.68~$\pm$~0.80&~~0.38&~1&\cr
ah   &01:03:16.05&-70:51:22.6&~30.00&162.74&47868.5455&217.46~$\pm$~1.23&&&&\cr
&&&&&47874.6135&215.75~$\pm$~1.33&216.67~$\pm$~0.90&~~0.89&~1&\cr
aj   &01:03:16.76&-70:51:11.6&~21.57&144.92&47868.5474&232.15~$\pm$~1.15&&&&\cr
ak   &01:03:11.06&-70:51:02.1&~17.64&242.48&47868.5541&227.15~$\pm$~1.12&&&&\cr
am   &01:03:10.30&-70:51:06.3&~22.99&237.49&47868.5568&228.86~$\pm$~1.25&&&&\cr
an   &01:03:11.86&-70:51:02.6&~14.56&233.55&47868.5586&223.35~$\pm$~0.97&&&&\cr
ap   &01:03:12.63&-70:50:50.0&~~8.85&296.49&47868.5633&223.59~$\pm$~1.43&&&&\cr
aq   &01:03:13.46&-70:50:50.1&~~5.44&315.08&47868.5651&217.22~$\pm$~1.26&&&&\cr
ar   &01:03:13.48&-70:50:52.9&~~3.88&285.68&47868.5668&218.86~$\pm$~1.42&&&&\cr
as   &01:03:14.08&-70:50:48.3&~~5.70&352.07&47868.5688&224.34~$\pm$~1.08&&&&\cr
at   &01:03:12.56&-70:50:37.9&~18.05&332.74&47868.5704&221.94~$\pm$~0.62&&&&\cr
&&&&&47873.6277&222.98~$\pm$~0.82&222.32~$\pm$~0.49&~~1.02&~1&\cr
au   &01:03:12.95&-70:50:35.3&~19.70&341.20&47868.5721&208.60~$\pm$~0.92&&&&\cr
av   &01:03:14.94&-70:51:00.1&~~7.05&150.75&47868.5743&229.62~$\pm$~0.68&&&&\cr
aw   &01:03:15.93&-70:50:54.0&~~8.32&~~90.3&47868.5762&219.67~$\pm$~1.01&&&&\cr
ax   &01:03:19.16&-70:50:59.5&~24.84&102.92&47868.5781&228.20~$\pm$~1.05&&&&\cr
ay   &01:03:19.00&-70:50:51.2&~23.59&~~83.3&47868.5797&222.15~$\pm$~0.71&&&&\cr
az   &01:03:17.55&-70:50:41.8&~20.32&~~53.2&47868.5816&224.64~$\pm$~1.32&&&&\cr
b    &01:03:13.77&-70:51:09.1&~15.33&188.68&47867.5307&208.12~$\pm$~0.61&&&&\cr
ba   &01:03:19.32&-70:50:34.0&~31.99&~~51.4&47869.5228&233.67~$\pm$~0.62&&&&\cr
bb   &01:03:20.62&-70:50:29.8&~39.61&~~52.4&47869.5287&238.13~$\pm$~0.71&&&&\cr
&&&&&48607.5479&239.66~$\pm$~0.99&238.65~$\pm$~0.58&~~1.58&~1&\cr
bc   &01:03:16.44&-70:50:26.9&~29.14&~~21.8&47869.5304&221.21~$\pm$~1.53&&&&\cr
\spc
\baselineskip=12pt
\endinsert
\vfill\dosupereject

\pageinsert
\baselineskip=10pt
\eightpoint\rm
\tabh{\taonen}{Radial Velocities}
\span\tcoli && \span\tcol{0.0} \cr
\topperc

Star & $\alpha$ & $\delta$ & R & PA & HJD & $v_r$ & $\overline{v_r}$ & $\chi^2$
& $\nu$ & Comments \cr
& (2000.0) & (2000.0) & (arcsec) & ($^{\circ}$) & -2400000 & (km\ s$^{-1}$) &
(\kms) \cr
\spacer

bd   &01:03:14.99&-70:50:26.8&~27.40&~~7.74&47869.5321&233.03~$\pm$~1.12&&&&\cr
be   &01:03:14.77&-70:50:14.5&~39.54&~~3.78&47869.5339&220.51~$\pm$~1.27&&&&\cr
bf   &01:03:12.90&-70:50:07.7&~46.72&351.88&47869.5357&224.95~$\pm$~0.97&&&&\cr
bg   &01:03:11.54&-70:50:22.7&~33.96&336.96&47869.5375&224.82~$\pm$~0.91&&&&\cr
bh   &01:03:10.24&-70:50:15.5&~43.20&332.88&47869.5392&226.67~$\pm$~0.79&&&&\cr
bj   &01:03:09.03&-70:50:20.0&~42.55&322.92&47869.5409&217.02~$\pm$~0.70&&&&\cr
bk   &01:03:09.99&-70:50:21.8&~38.36&326.94&47869.5426&226.17~$\pm$~1.20&&&&\cr
bm   &01:03:05.55&-70:50:30.9&~48.59&298.30&47869.5461&225.73~$\pm$~0.84&&&&\cr
bn   &01:03:05.43&-70:50:28.2&~50.43&300.69&47869.5478&224.88~$\pm$~1.19&&&&\cr
bp   &01:03:06.05&-70:50:36.6&~43.88&293.27&47869.5540&217.38~$\pm$~1.67&&&&\cr
bq   &01:03:06.69&-70:50:35.5&~41.49&296.39&47869.5557&225.86~$\pm$~1.77&&&&\cr
br   &01:03:07.22&-70:50:48.5&~34.97&278.95&47869.5575&225.00~$\pm$~1.33&&&&\cr
bs   &01:03:09.88&-70:50:48.2&~22.21&284.99&47869.5592&224.57~$\pm$~2.15&&&&\cr
bt   &01:03:05.65&-70:50:59.0&~42.57&263.17&47869.5627&223.41~$\pm$~0.71&&&&\cr
bu   &01:03:04.78&-70:50:58.7&~46.79&264.16&47869.5643&228.49~$\pm$~0.87&&&&\cr
bv   &01:03:06.06&-70:51:09.2&~43.04&249.23&47869.5661&222.72~$\pm$~0.96&&&&\cr
bw   &01:03:09.62&-70:51:00.4&~23.63&254.15&47869.5680&222.17~$\pm$~2.97&&&&\cr
bx   &01:03:11.69&-70:50:51.1&~12.87&282.79&47869.5698&222.19~$\pm$~1.35&&&&\cr
by   &01:03:09.15&-70:51:16.7&~33.83&227.74&47869.5715&236.62~$\pm$~0.60&&&&\cr
&&&&&47873.5612&234.79~$\pm$~0.80&235.96~$\pm$~0.48&~~3.35&~1&\cr
bz   &01:03:11.65&-70:51:25.9&~34.40&201.74&47869.5733&228.70~$\pm$~1.93&&&&\cr
c    &01:03:13.00&-70:51:14.3&~21.25&196.69&47867.5325&209.72~$\pm$~1.14&&&&\cr
ca   &01:03:11.57&-70:51:37.1&~45.11&196.93&47869.5750&231.81~$\pm$~0.91&&&&\cr
cb   &01:03:10.19&-70:51:35.1&~45.72&205.83&47869.5767&216.34~$\pm$~2.55&&&&\cr
cc   &01:03:12.15&-70:51:41.3&~48.45&192.25&47869.5785&221.65~$\pm$~0.80&&&&\cr
cd   &01:03:13.26&-70:51:33.4&~39.74&186.97&47869.5804&222.85~$\pm$~1.23&&&&\cr
ce   &01:03:07.80&-70:50:11.3&~53.14&323.36&47870.5276&220.39~$\pm$~1.23&&&&\cr
cf   &01:03:07.37&-70:50:11.8&~54.04&321.25&47870.5293&212.28~$\pm$~0.96&&&&\cr
&&&&&48607.5436&212.15~$\pm$~1.10&212.22~$\pm$~0.72&~~0.01&~1&\cr
cg   &01:03:09.07&-70:50:53.4&~25.45&271.23&47870.5310&218.64~$\pm$~0.97&&&&\cr
ch   &01:03:12.24&-70:51:33.5&~40.76&193.97&47870.5347&221.22~$\pm$~1.16&&&&\cr
cj   &01:03:17.56&-70:51:18.4&~29.40&146.26&47870.5365&220.90~$\pm$~1.98&&&&\cr
ck   &01:03:16.53&-70:51:02.1&~13.91&125.88&47870.5399&223.56~$\pm$~1.53&&&&\cr
cm   &01:03:18.84&-70:51:14.8&~30.77&132.66&47870.5418&224.94~$\pm$~0.97&&&&\cr
&&&&&47874.6171&223.80~$\pm$~1.06&224.42~$\pm$~0.72&~~0.63&~1&\cr
cn   &01:03:18.91&-70:51:26.1&~39.52&144.46&47870.5436&231.95~$\pm$~1.26&&&&\cr
&&&&&47870.5493&232.53~$\pm$~1.10&232.28~$\pm$~0.83&~~0.12&~1&\cr
cp   &01:03:17.47&-70:51:30.0&~39.40&156.22&47870.5455&221.31~$\pm$~1.91&&&&\cr
cq   &01:03:15.96&-70:51:36.3&~43.19&168.70&47870.5474&227.15~$\pm$~1.63&&&&\cr
cr   &01:03:20.83&-70:51:22.0&~42.87&130.88&47870.5513&220.32~$\pm$~0.89&&&&\cr
&&&&&47874.6153&220.43~$\pm$~1.07&220.36~$\pm$~0.68&~~0.01&~1&\cr
cs   &01:03:21.35&-70:51:20.7&~44.04&127.42&47870.5577&219.62~$\pm$~1.21&&&&\cr
ct   &01:03:21.80&-70:51:27.0&~49.76&131.64&47870.5599&219.06~$\pm$~1.30&&&&\cr
cu   &01:03:23.55&-70:51:21.0&~53.20&120.58&47870.5618&205.17~$\pm$~1.16&&&&\cr
&&&&&48605.6023&206.70~$\pm$~1.57&205.71~$\pm$~0.93&~~0.61&~1&\cr
cv   &01:03:23.65&-70:51:08.5&~48.53&107.46&47870.5637&223.82~$\pm$~0.74&&&&\cr
cw   &01:03:24.61&-70:51:08.9&~53.17&106.35&47870.5656&222.75~$\pm$~1.37&&&&\cr
cx   &01:03:24.64&-70:51:01.1&~51.67&~~97.9&47870.5676&220.97~$\pm$~2.75&&&&\cr
\spc
\baselineskip=12pt
\endinsert
\vfill\dosupereject

\pageinsert
\baselineskip=10pt
\eightpoint\rm
\tabh{\taonen}{Radial Velocities}
\span\tcoli && \span\tcol{0.0} \cr
\topperc

Star & $\alpha$ & $\delta$ & R & PA & HJD & $v_r$ & $\overline{v_r}$ & $\chi^2$
& $\nu$ & Comments \cr
& (2000.0) & (2000.0) & (arcsec) & ($^{\circ}$) & -2400000 & (km\ s$^{-1}$) &
(\kms) \cr
\spacer

cy   &01:03:23.92&-70:50:45.1&~48.45&~~79.5&47870.5694&219.24~$\pm$~1.43&&&&\cr
cz   &01:03:21.33&-70:50:49.7&~35.15&~~83.0&47870.5715&216.59~$\pm$~2.48&&&&\cr
d    &01:03:12.61&-70:51:02.9&~12.02&221.86&47867.5349&214.11~$\pm$~0.79&&&&\cr
da   &01:03:18.70&-70:50:32.8&~30.48&~~46.0&47870.5735&221.44~$\pm$~1.27&&&&\cr
db   &01:03:17.34&-70:50:26.3&~31.58&~~28.9&47870.5752&216.13~$\pm$~1.52&&&&\cr
dc   &01:03:23.91&-70:50:24.5&~55.97&~~58.2&47870.5770&221.63~$\pm$~1.02&&&&\cr
de   &01:03:24.31&-70:50:19.2&~60.53&~~54.9&47870.5787&221.46~$\pm$~1.42&&&&\cr
&&&&&47871.5485&223.78~$\pm$~1.13&222.88~$\pm$~0.88&~~1.63&~1&\cr
df   &01:03:21.70&-70:50:43.3&~38.23&~~73.8&47870.5808&226.60~$\pm$~2.31&&&&\cr
dg   &01:03:16.38&-70:50:04.7&~50.36&~~12.0&47870.5831&224.72~$\pm$~1.25&&&&\cr
dh   &01:03:06.46&-70:51:40.0&~59.88&219.72&47870.5849&225.04~$\pm$~1.07&&&&\cr
dj   &01:03:03.89&-70:51:31.9&~63.51&233.28&47870.5867&220.39~$\pm$~0.68&&&&\cr
dk   &01:03:25.78&-70:49:47.9&~87.12&~~40.7&47871.5289&207.16~$\pm$~2.03&&&&\cr
&&&&&48605.6207&212.71~$\pm$~1.93&210.08~$\pm$~1.40&~~3.93&~1&\cr
dm   &01:03:25.19&-70:49:58.9&~77.05&~~44.4&47871.5310&224.78~$\pm$~0.84&&&&\cr
dn   &01:03:23.13&-70:49:54.6&~73.74&~~36.4&47871.5328&223.72~$\pm$~2.03&&&&\cr
dp   &01:03:20.01&-70:49:55.2&~65.26&~~25.8&47871.5344&230.01~$\pm$~1.47&&&&\cr
dq   &01:03:20.54&-70:49:36.8&~83.15&~~21.9&47871.5364&217.51~$\pm$~2.19&&&&\cr
dr   &01:03:21.97&-70:49:58.2&~67.50&~~34.3&47871.5389&223.41~$\pm$~1.98&&&&\cr
ds   &01:03:25.00&-70:50:08.5&~69.79&~~49.3&47871.5411&220.58~$\pm$~2.07&&&&\cr
dt   &01:03:28.37&-70:50:40.8&~70.77&~~79.3&47871.5549&222.70~$\pm$~1.32&&&&\cr
du   &01:03:11.09&-70:49:43.4&~72.23&347.60&47872.5686&221.13~$\pm$~1.26&&&&\cr
dv   &01:03:07.76&-70:49:46.2&~74.89&334.77&47872.5704&234.26~$\pm$~0.69&&&&\cr
&&&&&48605.6518&236.18~$\pm$~1.02&234.86~$\pm$~0.57&~~2.43&~1&\cr
dw   &01:03:03.15&-70:50:30.2&~59.53&293.49&47872.5722&227.12~$\pm$~0.71&&&&\cr
dx   &01:03:02.78&-70:50:16.9&~67.49&303.27&47872.5739&216.03~$\pm$~0.93&&&&\cr
dy   &01:03:01.43&-70:49:59.8&~83.12&310.63&47872.5756&220.18~$\pm$~0.95&&&&\cr
&&&&&47873.5629&220.28~$\pm$~0.62&220.25~$\pm$~0.52&~~0.01&~1&\cr
dz   &01:03:01.12&-70:50:19.3&~73.29&298.19&47872.5773&212.94~$\pm$~2.01&&&&\cr
&&&&&48605.6535&213.16~$\pm$~1.54&213.08~$\pm$~1.22&~~0.01&~1&\cr
e    &01:03:10.62&-70:50:53.8&~17.81&270.47&47867.5367&233.57~$\pm$~0.72&&&&\cr
&&&&&47873.6263&233.17~$\pm$~0.80&233.39~$\pm$~0.54&~~0.14&~1&\cr
ea   &01:03:01.92&-70:51:17.5&~65.03&248.74&47873.5646&235.29~$\pm$~1.68&&&&\cr
&&&&&48605.6585&238.98~$\pm$~1.48&237.37~$\pm$~1.11&~~2.72&~1&\cr
eb   &01:03:00.13&-70:51:00.2&~69.71&264.83&47873.5669&217.24~$\pm$~1.62&&&&\cr
ec   &01:02:57.93&-70:50:40.4&~81.41&279.55&47873.5694&225.53~$\pm$~1.64&&&&\cr
ed   &01:03:08.91&-70:51:49.6&~61.52&205.22&47873.6182&226.04~$\pm$~0.89&&&&\cr
edd  &01:03:15.41&-70:51:42.6&~48.99&173.26&47874.5444&223.69~$\pm$~1.01&&&&\cr
ef   &01:03:10.94&-70:51:54.2&~62.40&195.07&47873.6197&215.07~$\pm$~0.93&&&&\cr
&&&&&48607.5690&215.01~$\pm$~0.78&215.03~$\pm$~0.60&~~0.00&~1&\cr
eff  &01:03:14.26&-70:51:58.0&~64.05&179.91&47874.5461&220.14~$\pm$~1.32&&&&\cr
eg   &01:03:19.30&-70:52:26.6&~95.93&164.98&47874.5479&214.82~$\pm$~0.93&&&&\cr
&&&&&48607.5636&215.14~$\pm$~0.85&214.99~$\pm$~0.63&~~0.06&~1&\cr
eh   &01:03:21.90&-70:52:04.0&~79.54&151.74&47874.5497&228.24~$\pm$~0.96&&&&\cr
ej   &01:03:23.65&-70:52:01.4&~81.80&145.56&47874.5513&234.67~$\pm$~0.77&&&&\cr
ek   &01:03:23.77&-70:51:49.8&~72.92&140.01&47874.5530&224.68~$\pm$~1.37&&&&\cr
em   &01:03:24.80&-70:51:51.5&~77.53&137.95&47874.5547&239.52~$\pm$~0.78&&&&\cr
&&&&&48605.6049&237.87~$\pm$~1.03&238.92~$\pm$~0.62&~~1.63&~1&\cr
\spc
\baselineskip=12pt
\endinsert
\vfill\dosupereject

\pageinsert
\baselineskip=10pt
\eightpoint\rm
\tabh{\taonen}{Radial Velocities}
\span\tcoli && \span\tcol{0.0} \cr
\topperc

Star & $\alpha$ & $\delta$ & R & PA & HJD & $v_r$ & $\overline{v_r}$ & $\chi^2$
& $\nu$ & Comments \cr
& (2000.0) & (2000.0) & (arcsec) & ($^{\circ}$) & -2400000 & (km\ s$^{-1}$) &
(\kms) \cr
\spacer

en   &01:03:25.95&-70:52:03.8&~90.53&140.51&47874.5565&238.43~$\pm$~1.11&&&&\cr
&&&&&48605.6073&236.81~$\pm$~1.12&237.63~$\pm$~0.79&~~1.06&~1&\cr
ep   &01:03:28.45&-70:51:29.4&~78.39&116.92&47874.5650&227.74~$\pm$~0.96&&&&\cr
eq   &01:03:25.59&-70:51:14.3&~59.44&110.04&47874.5716&223.95~$\pm$~2.08&&&&\cr
er   &01:03:31.94&-70:51:15.8&~89.79&104.12&47874.5741&237.84~$\pm$~2.58&&&&\cr
&&&&&48605.6107&234.44~$\pm$~1.79&235.54~$\pm$~1.47&~~1.17&~1&\cr
es   &01:03:18.36&-70:51:50.3&~59.88&160.23&47874.6101&217.60~$\pm$~1.87&&&&\cr
et   &01:03:17.02&-70:51:43.9&~51.79&164.69&47874.6118&211.16~$\pm$~1.53&&&&\cr
&&&&&48607.5666&210.09~$\pm$~1.34&210.55~$\pm$~1.01&~~0.28&~1&\cr
eu   &01:03:13.64&-70:50:58.7&~~5.59&211.86&47874.6190&232.54~$\pm$~0.98&&&&\cr
ev   &01:03:12.46&-70:50:59.8&~10.53&236.26&47874.6205&230.00~$\pm$~0.88&&&&\cr
ew   &01:03:10.83&-70:50:58.5&~17.39&254.82&47874.6222&231.64~$\pm$~1.15&&&&\cr
f    &01:03:11.19&-70:50:46.3&~16.85&297.00&47867.5387&230.67~$\pm$~0.93&&&&\cr
&&&&&47868.5606&232.80~$\pm$~1.15&&&&\cr
&&&&&47869.5609&233.23~$\pm$~0.91&&&&\cr
&&&&&47870.5329&230.69~$\pm$~1.11&&&&\cr
&&&&&47872.5794&232.42~$\pm$~1.06&&&&\cr
&&&&&47873.6385&232.65~$\pm$~1.20&&&&\cr
&&&&&47874.6239&233.12~$\pm$~0.97&&&&\cr
&&&&&48605.5998&232.95~$\pm$~1.07&&&&\cr
&&&&&48605.6732&231.12~$\pm$~1.20&&&&\cr
&&&&&48607.5389&231.67~$\pm$~1.40&232.18~$\pm$~0.34&~~8.63&~9&\cr
g    &01:03:08.88&-70:50:40.0&~29.84&297.86&47867.5405&214.15~$\pm$~0.94&&&&\cr
h    &01:03:10.21&-70:50:36.3&~26.55&311.66&47867.5422&219.09~$\pm$~0.94&&&&\cr
&&&&&47869.5444&220.12~$\pm$~0.75&219.72~$\pm$~0.59&~~0.73&~1&\cr
j    &01:03:10.68&-70:50:32.5&~27.70&320.75&47867.5441&231.75~$\pm$~0.78&&&&\cr
k    &01:03:10.46&-70:50:30.0&~30.33&322.15&47867.5459&225.17~$\pm$~1.08&&&&\cr
m    &01:03:11.39&-70:50:37.6&~21.54&319.37&47867.5531&222.99~$\pm$~1.07&&&&\cr
n    &01:03:13.94&-70:50:40.1&~13.93&353.91&47867.5550&216.43~$\pm$~1.35&&&&\cr
p    &01:03:13.60&-70:50:36.7&~17.54&349.65&47867.5568&228.24~$\pm$~0.94&&&&\cr
&&&&&47873.6292&229.18~$\pm$~1.08&228.65~$\pm$~0.71&~~0.43&~1&\cr
q    &01:03:13.68&-70:50:31.8&~22.32&352.91&47867.5585&227.56~$\pm$~0.80&&&&\cr
r    &01:03:15.06&-70:50:32.1&~22.22&~~10.4&47867.5603&217.66~$\pm$~0.97&&&&\cr
&&&&&47873.6315&219.93~$\pm$~1.09&218.66~$\pm$~0.72&~~2.42&~1&\cr
s    &01:03:15.99&-70:50:43.5&~13.54&~~39.5&47867.5623&222.53~$\pm$~0.80&&&&\cr
&&&&&47873.6335&222.08~$\pm$~0.87&222.32~$\pm$~0.59&~~0.14&~1&\cr
t    &01:03:14.06&-70:50:51.2&~~2.89&342.15&47867.5645&225.73~$\pm$~1.18&&&&\cr
u    &01:03:15.19&-70:51:03.4&~10.54&153.68&47867.5664&232.50~$\pm$~0.86&&&&\cr
v    &01:03:15.55&-70:50:59.6&~~8.57&131.24&47867.5685&214.04~$\pm$~1.31&&&&\cr
w    &01:03:20.04&-70:50:54.8&~28.55&~~91.7&47867.5702&215.61~$\pm$~0.67&&&&\cr
x    &01:03:17.41&-70:50:54.8&~15.62&~~93.1&47868.5255&227.39~$\pm$~0.82&&&&\cr
&&&&&47873.6353&226.56~$\pm$~1.32&227.16~$\pm$~0.70&~~0.29&~1&\cr
y    &01:03:17.70&-70:51:05.4&~20.52&123.93&47868.5273&219.28~$\pm$~0.71&&&&\cr
&&&&&47873.6368&218.91~$\pm$~0.81&219.12~$\pm$~0.53&~~0.12&~1&\cr
z    &01:03:16.42&-70:51:07.1&~16.97&140.80&47868.5291&215.14~$\pm$~1.37&&&&\cr
&&&&&47870.5382&215.51~$\pm$~0.88&215.40~$\pm$~0.74&~~0.05&~1&\cr
\sp
\baselineskip=12pt
\endinsert
\vfill\dosupereject

\pageinsert
\tabh{\taone}{Expected Fraction of Velocity Differences $\ge$ (6, 8) \kms}
\span\tcoli && \span\tcol{0.3} \cr
\topper

\hfil $x_b$ & $e = 0$ & $e = f(e)$ \cr
\spacer

0.00 & (0.001, 0.000)  & (0.001, 0.000) \cr
0.05 & (0.022, 0.018)  & (0.014, 0.001) \cr
0.10 & (0.047, 0.041)  & (0.024, 0.019) \cr
0.15 & (0.071, 0.063)  & (0.039, 0.031) \cr
0.20 & (0.096, 0.085)  & (0.049, 0.040) \cr
0.25 & (0.121, 0.108)  & (0.069, 0.055) \cr
0.30 & (0.147, 0.130)  & (0.078, 0.065) \cr
0.35 & (0.171, 0.150)  & (0.096, 0.079) \cr
0.40 & (0.199, 0.177)  & (0.107, 0.090) \cr
0.45 & (0.221, 0.195)  & (0.132, 0.111) \cr
0.50 & (0.249, 0.224)  & (0.139, 0.118) \cr
\sp
\vfill
\endinsert
\vfill\dosupereject

\pageinsert
\tabh{\taone}{Mass Bins}
\span\tcoli && \span\tcol{.3} \cr
\topper

\hfil Bin & m$_{min}$ & m$_{max}$ & Contents \cr
\hfil  & (M\sol) &  (M\sol) \cr
\spacer

{}~1 & 0.15 & 0.20 & MS \cr
{}~2 & 0.20 & 0.25 & MS \cr
{}~3 & 0.25 & 0.30 & MS \cr
{}~4 & 0.30 & 0.35 & MS \cr
{}~5 & 0.35 & 0.40 & MS \cr
{}~6 & 0.40 & 0.45 & MS \cr
{}~7 & 0.45 & 0.50 & MS \cr
{}~8 & 0.50 & 0.55 & MS \& WD \cr
{}~9 & 0.55 & 0.60 & MS \cr
10 & 0.60 & 0.65 & MS \cr
11 & 0.65 & 0.70 & MS \& WD \cr
12 & 0.70 & 0.75 & MS \cr
13 & 0.75 & 0.80 & MS \cr
14 & 0.80 & 0.84 & MS \cr
15 & 0.84 & 0.87 & MS \& RG \& HB \cr
16 & 0.87 & 8.00 & WD \cr
\sp
\hskip1.27truein Notes:~~MS = main sequence stars

\hskip1.27truein \phantom{Notes:~~}WD = white dwarf stars

\hskip1.27truein \phantom{Notes:~~}RG = red giant stars

\hskip1.27truein \phantom{Notes:~~}HB = horizontal branch stars
\vfill
\endinsert
\vfill\dosupereject

\pageinsert
%remnants, cutoff=0.3
{
\tabhl{\taone}{King-Michie - Fitted Parameters - Model A}
\span\tcoli & \span\tcoll{.15}{8}{4} & \span\tcol{.15} & \span\tcol{.15} &
\span\tcol{.15} & \span\tcol{0.0}
& \span\tcoll{0.0}{8}{4} & \span\tcol{.15} & \span\tcol{0.0} & \span\tcol{0.0}
\cr
\topperl

 & & & \multispan3 Surface~Photometry & & \multispan3 Velocities \cr
\tablerule
\hfil r${_a}$ & x & W$_0$ & r$_s$ & c & $\chi_\nu^2$ & P($\chi_\nu^2$) & v$_s$
& $\zeta^2$ & P($|\zeta^2-N|$) \cr
\hfil (r$_s$) & &  & (pc) & & ($\nu=22$) & & (km s$^{-1}$) & & \cr
\tablerule

ISO & & $~7.5 \pm 0.1$ & $0.56 \pm 0.02$ & $~49 \pm ~4$ & 0.60 & 0.91 &$~8.9
\pm 0.5$ & 205.69 & 0.02 \cr
{}~10 & & $~7.4 \pm 0.1$ & $0.59 \pm 0.02$ & $~76 \pm 13$ & 0.51 & 0.96 & $~9.2
\pm 0.5$ & 205.99 & 0.07 \cr
{}~~5 & & $~6.8 \pm 0.1$ & $0.70 \pm 0.02$ & $148 \pmm{+67}{-51}$ & 1.02 & 0.37
& $~9.9 \pm 0.5$ & 206.55 & 0.26 \cr
{}~~3 & & $~6.0 \pm 0.1$ & $0.80 \pm 0.03$ & $133 \pmm{+85}{-47}$ & 2.28 & 0.00
& $11.2 \pm 0.6$ & 207.27 & 0.72 \cr
&&&&&&&&\cr
ISO & 0.0 & $~9.1 \pm 0.2$ & $0.36 \pm 0.02$ & $~56 \pm ~5$ & 1.83 & 0.01 &
$~8.5 \pm 0.4$ & 206.09 & 0.09 \cr
{}~10 & 0.0 & $~8.7 \pm 0.1$ & $0.39 \pm 0.02$ & $~87 \pm 14$ & 0.77 & 0.74 &
$~9.2 \pm 0.5$ & 206.69 & 0.22 \cr
{}~~5 & 0.0 & $~7.9 \pm 0.1$ & $0.50 \pm 0.02$ & $217 \pmm{+70}{-20}$ & 0.60 &
0.92 & $~9.9 \pm 0.5$ & 207.02 & 0.59 \cr
{}~~3 & 0.0 & $~6.7 \pm 0.1$ & $0.64 \pm 0.02$ & $173 \pmm{+63}{-47}$ & 1.38 &
0.08 & $11.1 \pm 0.6$ & 207.47 & 0.88 \cr
&&&&&&&&\cr
ISO & 0.5 & $10.0 \pm 0.2$ & $0.39 \pm 0.03$ & $~57 \pm ~5$ & 1.32 & 0.12 &
$~7.7 \pm 0.4$ & 205.54 & 0.01 \cr
{}~10 & 0.5 & $~9.6 \pm 0.2$ & $0.43 \pm 0.02$ & $~99 \pm 20$ & 0.64 & 0.86 &
$~8.4 \pm 0.4$ & 206.20 & 0.13 \cr
{}~~5 & 0.5 & $~8.4 \pm 0.1$ & $0.57 \pm 0.02$ & $192 \pmm{+70}{-50}$ & 0.93 &
0.49 & $~9.2 \pm 0.5$ & 206.63 & 0.34 \cr
{}~~3 & 0.5 & $~7.1 \pm 0.1$ & $0.71 \pm 0.02$ & $159 \pmm{+64}{-32}$ & 2.06 &
0.00 & $10.4 \pm 0.6$ & 207.09 & 0.66 \cr
&&&&&&&&\cr
ISO & 1.0 & $11.1 \pm 0.3$ & $0.48 \pm 0.03$ & $~59 \pm ~5$ & 0.66 & 0.85 &
$~7.0 \pm 0.4$ & 204.99 & 0.00 \cr
{}~10 & 1.0 & $10.9 \pm 0.2$ & $0.52 \pm 0.02$ & $212 \pm 60$ & 0.52 & 0.96 &
$~7.5 \pm 0.4$ & 205.39 & 0.01 \cr
{}~~5 & 1.0 & $~8.9 \pm 0.1$ & $0.70 \pm 0.02$ & $156 \pmm{+40}{-60}$ & 2.02 &
0.00 & $~8.7 \pm 0.4$ & 206.08 & 0.10 \cr
{}~~3 & 1.0 & $~7.4 \pm 0.1$ & $0.84 \pm 0.02$ & $140 \pmm{+61}{-46}$ & 3.51 &
0.00 & $~9.9 \pm 0.5$ & 206.59 & 0.36 \cr
&&&&&&&&\cr
ISO & 1.5 & $12.1 \pm 0.3$ & $0.58 \pm 0.03$ & $~72 \pm ~6$ & 0.60 & 0.91 &
$~6.7 \pm 0.3$ & 204.74 & 0.00 \cr
{}~10 & 1.5 & $10.9 \pm 0.2$ & $0.70 \pm 0.02$ & $151 \pmm{+30}{-50}$ & 1.28 &
0.16 & $~7.4 \pm 0.4$ & 205.01 & 0.00 \cr
{}~~5 & 1.5 & $~8.9 \pm 0.2$ & $0.87 \pm 0.02$ & $145 \pmm{+60}{-30}$ & 3.79 &
0.00 & $~8.5 \pm 0.4$ & 205.55 & 0.02 \cr
{}~~3 & 1.5 & $~7.4 \pm 0.1$ & $0.99 \pm 0.03$ & $128 \pmm{+56}{-26}$ & 5.55 &
0.00 & $~9.8 \pm 0.5$ & 206.12 & 0.15 \cr
&&&&&&&&\cr
ISO & 2.0 & $13.2 \pm 0.3$ & $0.67 \pm 0.02$ & $105 \pm 10$ & 0.99 & 0.41 &
$~6.7 \pm 0.3$ & 204.70 & 0.00 \cr
{}~10 & 2.0 & $10.7 \pm 0.3$ & $0.88 \pm 0.03$ & $136 \pm 40$ & 2.98 & 0.00 &
$~7.5 \pm 0.4$ & 204.83 & 0.00 \cr
{}~~5 & 2.0 & $~8.8 \pm 0.2$ & $1.02 \pm 0.03$ & $137 \pmm{+40}{-20}$ & 5.64 &
0.00 & $~8.6 \pm 0.4$ & 205.26 & 0.00 \cr
{}~~3 & 2.0 & $~7.2 \pm 0.1$ & $1.10 \pm 0.03$ & $~68 \pm 20$ & 7.68 & 0.00 &
$10.1 \pm 0.5$ & 206.08 & 0.11 \cr
\spl}
\endinsert
\vfill\dosupereject

\pageinsert
\baselineskip=12pt
\tabh{\taone}{King-Michie - Derived Parameters - Model A}
\span\tcoli && \span\tcol{.05} \cr
\topper

\hfil         & &                   &               &                   &
        & \multispan2 Population & \multispan2 Dynamic \cr
\hfil r$_a$ & x & $\mu_{V0}$ & L$_V$ & $\rho_{0}$ & M & (M/L$_V$)$_0$ &
(M/L$_V$) & (M/L$_V$)$_0$ & (M/L$_V$) \cr
\hfil (r$_s$) & & $10^4\choose{\hbox{L}_V\hbox{\sol} \hbox{pc}^{-3}}$ & $10^5$
L$_V$\sol & $10^4\choose{\hbox{M\sol\ pc}^{-3}}$ & (10$^5$M\sol) & \multispan2
\hfil (M/L$_V$)\sol\hfil & \multispan2 \hfil (M/L$_V$)\sol\hfil \cr
\spacer

ISO & & $3.0 \pm 0.2$ & $1.66 \pm 0.05$ & $4.2 \pm 0.5$ & $2.3 \pm 0.3$ & & &
$1.4 \pm 0.2$ & $1.4 \pm 0.2$ \cr
{}~10 & & $2.8 \pm 0.2$ & $1.72 \pm 0.06$ & $4.1 \pm 0.5$ & $2.5 \pm 0.3$ & & &
$1.4 \pm 0.2$ & $1.4 \pm 0.2$ \cr
{}~~5 & & $2.2 \pm 0.2$ & $1.73 \pm 0.06$ & $3.3 \pm 0.4$ & $2.7 \pm 0.3$ & & &
$1.6 \pm 0.2$ & $1.6 \pm 0.2$ \cr
{}~~3 & & $1.9 \pm 0.2$ & $1.65 \pm 0.05$ & $3.3 \pm 0.5$ & $2.9 \pm 0.3$ & & &
$1.7 \pm 0.2$ & $1.7 \pm 0.2$ \cr
\cr
ISO & 0.0 & $3.7 \pm 0.4$ & $1.57 \pm 0.04$ & $9.4 \pm 1.5$ & $2.3 \pm 0.2$ &
7.03 & 4.21 & $2.5 \pm 0.3$ & $1.5 \pm 0.2$ \cr
{}~10 & 0.0 & $3.5 \pm 0.3$ & $1.65 \pm 0.05$ & $9.0 \pm 1.0$ & $2.6 \pm 0.3$ &
6.88 & 4.21 & $2.6 \pm 0.3$ & $1.6 \pm 0.2$ \cr
{}~~5 & 0.0 & $2.6 \pm 0.3$ & $1.77 \pm 0.05$ & $6.4 \pm 1.0$ & $3.0 \pm 0.3$ &
6.03 & 4.21 & $2.5 \pm 0.3$ & $1.7 \pm 0.2$ \cr
{}~~3 & 0.0 & $2.1 \pm 0.2$ & $1.71 \pm 0.05$ & $5.1 \pm 0.6$ & $3.2 \pm 0.4$ &
5.44 & 4.21 & $2.4 \pm 0.3$ & $1.9 \pm 0.2$ \cr
\cr
ISO & 0.5 & $3.6 \pm 0.3$ & $1.60 \pm 0.04$ & $6.3 \pm 1.0$ & $2.6 \pm 0.3$ &
3.45 & 3.17 & $1.8 \pm 0.2$ & $1.6 \pm 0.2$ \cr
{}~10 & 0.5 & $3.2 \pm 0.2$ & $1.67 \pm 0.05$ & $6.2 \pm 1.0$ & $3.0 \pm 0.3$ &
3.38 & 3.17 & $1.9 \pm 0.2$ & $1.8 \pm 0.2$ \cr
{}~~5 & 0.5 & $2.3 \pm 0.2$ & $1.72 \pm 0.05$ & $4.4 \pm 0.5$ & $3.4 \pm 0.4$ &
2.98 & 3.17 & $1.9 \pm 0.2$ & $2.0 \pm 0.2$ \cr
{}~~3 & 0.5 & $1.9 \pm 0.2$ & $1.67 \pm 0.05$ & $3.6 \pm 0.4$ & $3.5 \pm 0.4$ &
2.77 & 3.17 & $1.9 \pm 0.2$ & $2.1 \pm 0.3$ \cr
\cr
ISO & 1.0 & $3.1 \pm 0.3$ & $1.65 \pm 0.05$ & $3.6 \pm 0.5$ & $3.2 \pm 0.4$ &
1.78 & 3.07 & $1.1 \pm 0.1$ & $2.0 \pm 0.2$ \cr
{}~10 & 1.0 & $2.6 \pm 0.2$ & $1.77 \pm 0.06$ & $3.4 \pm 0.4$ & $4.2 \pm 0.5$ &
1.66 & 3.07 & $1.3 \pm 0.1$ & $2.4 \pm 0.3$ \cr
{}~~5 & 1.0 & $2.0 \pm 0.1$ & $1.65 \pm 0.05$ & $2.6 \pm 0.3$ & $4.0 \pm 0.5$ &
1.63 & 3.07 & $1.3 \pm 0.1$ & $2.4 \pm 0.3$ \cr
{}~~3 & 1.0 & $1.7 \pm 0.1$ & $1.58 \pm 0.04$ & $2.4 \pm 0.3$ & $4.0 \pm 0.5$ &
1.62 & 3.07 & $1.4 \pm 0.2$ & $2.6 \pm 0.3$ \cr
\cr
ISO & 1.5 & $2.8 \pm 0.2$ & $1.71 \pm 0.05$ & $2.2 \pm 0.3$ & $4.6 \pm 0.5$ &
1.06 & 3.50 & $0.8 \pm 0.1$ & $2.7 \pm 0.3$ \cr
{}~10 & 1.5 & $2.1 \pm 0.1$ & $1.70 \pm 0.05$ & $1.8 \pm 0.2$ & $5.1 \pm 0.7$ &
1.01 & 3.50 & $0.9 \pm 0.1$ & $3.0 \pm 0.4$ \cr
{}~~5 & 1.5 & $1.7 \pm 0.1$ & $1.56 \pm 0.04$ & $1.6 \pm 0.2$ & $4.8 \pm 0.6$ &
1.07 & 3.50 & $1.0 \pm 0.1$ & $3.1 \pm 0.4$ \cr
{}~~3 & 1.5 & $1.5 \pm 0.1$ & $1.49 \pm 0.04$ & $1.6 \pm 0.2$ & $4.6 \pm 0.6$ &
1.16 & 3.50 & $1.1 \pm 0.1$ & $3.1 \pm 0.4$ \cr
\cr
ISO & 2.0 & $2.4 \pm 0.2$ & $1.83 \pm 0.06$ & $1.6 \pm 0.2$ & $7.8 \pm 1.0$ &
0.70 & 4.50 & $0.7 \pm 0.1$ & $4.3 \pm 0.5$ \cr
{}~10 & 2.0 & $1.7 \pm 0.1$ & $1.62 \pm 0.05$ & $1.2 \pm 0.1$ & $6.3 \pm 0.5$ &
0.79 & 4.50 & $0.7 \pm 0.1$ & $3.9 \pm 0.5$ \cr
{}~~5 & 2.0 & $1.5 \pm 0.1$ & $1.47 \pm 0.04$ & $1.2 \pm 0.1$ & $5.6 \pm 0.6$ &
0.92 & 4.50 & $0.8 \pm 0.1$ & $3.8 \pm 0.4$ \cr
{}~~3 & 2.0 & $1.6 \pm 0.1$ & $1.45 \pm 0.04$ & $1.4 \pm 0.2$ & $4.9 \pm 0.6$ &
1.14 & 4.50 & $0.9 \pm 0.1$ & $3.4 \pm 0.4$ \cr
\sp
\endinsert
\vfill\dosupereject

\pageinsert
%remnants, cutoff=0.3
{
\tabhl{\taone}{King-Michie - Fitted Parameters - Model B}
\span\tcoli & \span\tcoll{.15}{8}{4} & \span\tcol{.15} & \span\tcol{.15} &
\span\tcol{.15} & \span\tcol{0.0}
& \span\tcoll{0.0}{8}{4} & \span\tcol{.15} & \span\tcol{0.0} & \span\tcol{0.0}
\cr
\topperl

 & & & \multispan3 Surface~Photometry & & \multispan3 Velocities \cr
\tablerule
\hfil r${_a}$ & x & W$_0$ & r$_s$ & c & $\chi_\nu^2$ & P($\chi_\nu^2$) & v$_s$
& $\zeta^2$ & P($|\zeta^2-N|$) \cr
\hfil (r$_s$) & &  & (pc) & & ($\nu=22$) & & (km s$^{-1}$) & & \cr
\tablerule

ISO & 0.0 & $~9.1 \pm 0.2$ & $0.37 \pm 0.02$ & $~57 \pm ~5$ & 1.70 & 0.02 &
$~8.4 \pm 0.4$ & 206.01 & 0.08 \cr
{}~10 & 0.0 & $~8.7 \pm 0.1$ & $0.40 \pm 0.02$ & $~88 \pm 14$ & 0.73 & 0.79 &
$~9.1 \pm 0.5$ & 206.61 & 0.28 \cr
{}~~5 & 0.0 & $~7.9 \pm 0.1$ & $0.52 \pm 0.02$ & $221 \pmm{+70}{-20}$ & 0.63 &
0.90 & $~9.8 \pm 0.5$ & 206.96 & 0.55 \cr
{}~~3 & 0.0 & $~6.7 \pm 0.1$ & $0.65 \pm 0.02$ & $169 \pmm{+63}{-47}$ & 1.48 &
0.05 & $11.0 \pm 0.6$ & 207.44 & 0.86 \cr
&&&&&&&&\cr
ISO & 0.5 & $10.0 \pm 0.2$ & $0.40 \pm 0.03$ & $~58 \pm ~5$ & 1.20 & 0.19 &
$~7.6 \pm 0.4$ & 205.48 & 0.01 \cr
{}~10 & 0.5 & $~9.6 \pm 0.2$ & $0.45 \pm 0.02$ & $101 \pm 20$ & 0.61 & 0.90 &
$~8.3 \pm 0.4$ & 206.10 & 0.10 \cr
{}~~5 & 0.5 & $~8.4 \pm 0.1$ & $0.58 \pm 0.02$ & $184 \pmm{+70}{-50}$ & 1.03 &
0.36 & $~9.2 \pm 0.5$ & 206.58 & 0.31 \cr
{}~~3 & 0.5 & $~7.1 \pm 0.1$ & $0.72 \pm 0.02$ & $157 \pmm{+64}{-32}$ & 2.22 &
0.00 & $10.4 \pm 0.6$ & 207.04 & 0.62 \cr
&&&&&&&&\cr
ISO & 1.0 & $10.9 \pm 0.3$ & $0.50 \pm 0.03$ & $~59 \pm ~5$ & 0.58 & 0.93 &
$~7.0 \pm 0.4$ & 204.97 & 0.00 \cr
{}~10 & 1.0 & $10.8 \pm 0.2$ & $0.54 \pm 0.02$ & $195 \pm 60$ & 0.56 & 0.94 &
$~7.5 \pm 0.4$ & 205.37 & 0.00 \cr
{}~~5 & 1.0 & $~8.8 \pm 0.1$ & $0.72 \pm 0.02$ & $156 \pmm{+40}{-60}$ & 2.21 &
0.00 & $~8.6 \pm 0.4$ & 206.03 & 0.09 \cr
{}~~3 & 1.0 & $~7.4 \pm 0.1$ & $0.85 \pm 0.02$ & $145 \pmm{+61}{-46}$ & 3.73 &
0.00 & $~9.9 \pm 0.5$ & 206.55 & 0.34 \cr
&&&&&&&&\cr
ISO & 1.5 & $12.1 \pm 0.3$ & $0.58 \pm 0.03$ & $~75 \pm ~6$ & 0.60 & 0.91 &
$~6.7 \pm 0.3$ & 204.73 & 0.00 \cr
{}~10 & 1.5 & $10.8 \pm 0.2$ & $0.72 \pm 0.02$ & $156 \pmm{+30}{-50}$ & 1.42 &
0.08 & $~7.3 \pm 0.4$ & 204.99 & 0.00 \cr
{}~~5 & 1.5 & $~8.9 \pm 0.2$ & $0.89 \pm 0.02$ & $151 \pmm{+60}{-30}$ & 4.10 &
0.00 & $~8.4 \pm 0.4$ & 205.52 & 0.02 \cr
{}~~3 & 1.5 & $~7.4 \pm 0.1$ & $1.00 \pm 0.03$ & $127 \pmm{+56}{-26}$ & 5.93 &
0.00 & $~9.7 \pm 0.5$ & 206.12 & 0.15 \cr
&&&&&&&&\cr
ISO & 2.0 & $13.3 \pm 0.3$ & $0.67 \pm 0.02$ & $111 \pm 10$ & 0.94 & 0.48 &
$~6.6 \pm 0.3$ & 204.70 & 0.00 \cr
{}~10 & 2.0 & $10.6 \pm 0.3$ & $0.88 \pm 0.03$ & $129 \pm 40$ & 3.35 & 0.00 &
$~7.5 \pm 0.4$ & 204.83 & 0.00 \cr
{}~~5 & 2.0 & $~8.8 \pm 0.2$ & $1.03 \pm 0.03$ & $138 \pmm{+40}{-20}$ & 6.16 &
0.00 & $~8.6 \pm 0.4$ & 205.25 & 0.00 \cr
{}~~3 & 2.0 & $~7.4 \pm 0.1$ & $1.14 \pm 0.03$ & $135 \pmm{+40}{-20}$ & 8.30 &
0.00 & $~9.8 \pm 0.5$ & 205.82 & 0.06 \cr
\spl}
\endinsert
\vfill\dosupereject

\pageinsert
\baselineskip=12pt
\tabh{\taone}{King-Michie - Derived Parameters - Model B}
\span\tcoli && \span\tcol{.05} \cr
\topper

\hfil         & &                   &               &                   &
        & \multispan2 Population & \multispan2 Dynamic \cr
\hfil r$_a$ & x & $\mu_{V0}$ & L$_V$ & $\rho_{0}$ & M & (M/L$_V$)$_0$ &
(M/L$_V$) & (M/L$_V$)$_0$ & (M/L$_V$) \cr
\hfil (r$_s$) & & $10^4\choose{\hbox{L}_V\hbox{\sol} \hbox{pc}^{-3}}$ & $10^5$
L$_V$\sol & $10^4\choose{\hbox{M\sol\ pc}^{-3}}$ & (10$^5$M\sol) & \multispan2
\hfil (M/L$_V$)\sol\hfil & \multispan2 \hfil (M/L$_V$)\sol\hfil \cr
\spacer

ISO & 0.0 & $3.7 \pm 0.4$ & $1.58 \pm 0.04$ & $8.8 \pm 1.5$ & $2.4 \pm 0.2$ &
4.06 & 2.56 & $2.4 \pm 0.3$ & $1.5 \pm 0.2$ \cr
{}~10 & 0.0 & $3.5 \pm 0.3$ & $1.66 \pm 0.05$ & $8.5 \pm 1.0$ & $2.6 \pm 0.3$ &
3.97 & 2.56 & $2.4 \pm 0.3$ & $1.6 \pm 0.2$ \cr
{}~~5 & 0.0 & $2.6 \pm 0.3$ & $1.77 \pm 0.05$ & $6.0 \pm 1.0$ & $3.0 \pm 0.3$ &
3.47 & 2.56 & $2.4 \pm 0.3$ & $1.7 \pm 0.2$ \cr
{}~~3 & 0.0 & $2.1 \pm 0.2$ & $1.70 \pm 0.05$ & $4.8 \pm 0.6$ & $3.2 \pm 0.4$ &
3.16 & 2.56 & $2.3 \pm 0.3$ & $1.9 \pm 0.2$ \cr
\cr
ISO & 0.5 & $3.6 \pm 0.3$ & $1.61 \pm 0.04$ & $6.0 \pm 1.0$ & $2.6 \pm 0.3$ &
2.01 & 1.96 & $1.7 \pm 0.2$ & $1.6 \pm 0.2$ \cr
{}~10 & 0.5 & $3.2 \pm 0.2$ & $1.68 \pm 0.05$ & $5.7 \pm 1.0$ & $3.0 \pm 0.3$ &
1.96 & 1.96 & $1.8 \pm 0.2$ & $1.8 \pm 0.2$ \cr
{}~~5 & 0.5 & $2.3 \pm 0.2$ & $1.71 \pm 0.05$ & $4.1 \pm 0.5$ & $3.4 \pm 0.4$ &
1.75 & 1.96 & $1.8 \pm 0.2$ & $2.0 \pm 0.2$ \cr
{}~~3 & 0.5 & $1.9 \pm 0.2$ & $1.66 \pm 0.05$ & $3.4 \pm 0.4$ & $3.5 \pm 0.4$ &
1.64 & 1.96 & $1.8 \pm 0.2$ & $2.1 \pm 0.3$ \cr
\cr
ISO & 1.0 & $3.0 \pm 0.3$ & $1.66 \pm 0.05$ & $3.3 \pm 0.5$ & $3.3 \pm 0.4$ &
1.04 & 1.91 & $1.1 \pm 0.1$ & $2.0 \pm 0.2$ \cr
{}~10 & 1.0 & $2.6 \pm 0.2$ & $1.75 \pm 0.06$ & $3.2 \pm 0.4$ & $4.2 \pm 0.5$ &
1.00 & 1.91 & $1.2 \pm 0.1$ & $2.4 \pm 0.3$ \cr
{}~~5 & 1.0 & $1.9 \pm 0.1$ & $1.64 \pm 0.05$ & $2.4 \pm 0.3$ & $4.0 \pm 0.5$ &
0.97 & 1.91 & $1.3 \pm 0.1$ & $2.4 \pm 0.3$ \cr
{}~~3 & 1.0 & $1.7 \pm 0.1$ & $1.57 \pm 0.04$ & $2.2 \pm 0.3$ & $4.0 \pm 0.5$ &
0.97 & 1.91 & $1.3 \pm 0.2$ & $2.6 \pm 0.3$ \cr
\cr
ISO & 1.5 & $2.8 \pm 0.2$ & $1.72 \pm 0.05$ & $2.2 \pm 0.3$ & $4.7 \pm 0.5$ &
0.65 & 2.22 & $0.8 \pm 0.1$ & $2.7 \pm 0.3$ \cr
{}~10 & 1.5 & $2.0 \pm 0.1$ & $1.69 \pm 0.05$ & $1.7 \pm 0.2$ & $5.1 \pm 0.7$ &
0.61 & 2.22 & $0.9 \pm 0.1$ & $3.0 \pm 0.4$ \cr
{}~~5 & 1.5 & $1.6 \pm 0.1$ & $1.54 \pm 0.04$ & $1.5 \pm 0.2$ & $4.8 \pm 0.6$ &
0.65 & 2.22 & $0.9 \pm 0.1$ & $3.1 \pm 0.4$ \cr
{}~~3 & 1.5 & $1.5 \pm 0.1$ & $1.47 \pm 0.04$ & $1.6 \pm 0.2$ & $4.6 \pm 0.6$ &
0.73 & 2.22 & $1.0 \pm 0.1$ & $3.1 \pm 0.4$ \cr
\cr
ISO & 2.0 & $2.4 \pm 0.2$ & $1.82 \pm 0.06$ & $1.6 \pm 0.2$ & $8.1 \pm 1.0$ &
0.43 & 2.90 & $0.7 \pm 0.1$ & $4.5 \pm 0.5$ \cr
{}~10 & 2.0 & $1.7 \pm 0.1$ & $1.59 \pm 0.05$ & $1.2 \pm 0.1$ & $6.2 \pm 0.5$ &
0.50 & 2.90 & $0.7 \pm 0.1$ & $3.9 \pm 0.5$ \cr
{}~~5 & 2.0 & $1.5 \pm 0.1$ & $1.44 \pm 0.04$ & $1.2 \pm 0.1$ & $5.5 \pm 0.6$ &
0.57 & 2.90 & $0.8 \pm 0.1$ & $3.8 \pm 0.4$ \cr
{}~~3 & 2.0 & $1.4 \pm 0.1$ & $1.36 \pm 0.04$ & $1.2 \pm 0.2$ & $5.3 \pm 0.6$ &
0.67 & 2.90 & $0.9 \pm 0.1$ & $3.9 \pm 0.4$ \cr
\sp
\endinsert
\vfill\dosupereject

\bye